\documentclass[algorithms,article,accept,moreauthors,pdftex]{Definitions/mdpi} 
\firstpage{1} 
\pubvolume{14}
\issuenum{1}
\articlenumber{344}
\pubyear{2021}
\copyrightyear{2021}
\externaleditor{Academic Editor: {Frank} Werner} % To Editors: Please add.
\datereceived{31 October 2021} 
\dateaccepted{24 November 2021} 
\datepublished{27 November 2021} 
\hreflink{https://doi.org/10.3390/a14120344} 
\usepackage[ruled]{algorithm2e}  

\usepackage[caption=false,font=footnotesize,labelfont=sf,textfont=sf]{subfig}
\usepackage{booktabs}
\usepackage{xcolor}
\usepackage{float}
\Title{A Visual Mining Approach to Improved Multiple-Instance~Learning}
\TitleCitation{A Visual Mining Approach to Improved Multiple-Instance Learning}
\Author{Sonia Castelo $^{1,2}$, Moacir Ponti $^{3}$ and Rosane Minghim $^{4,}$*}
\AuthorNames{Sonia Castelo, Moacir Ponti and Rosane Minghim}
\AuthorCitation{Castelo, S.; Ponti, M.; Minghim, R.}
\address{%MDPI: Please complate the addresses (Including the department/school/faculty/campus, University/Institute/Company, we added the cities and postcodes, please confirm. %authors: Added and Confirmed.
$^{1}$ \quad EPIS, Universidad Nacional de San Agustín de Arequipa, Arequipa 04001, Peru\\
$^{2}$ \quad {NYU Tandon School of Engineering, New York University, Brooklyn, NY 11201, USA}; s.castelo@nyu.edu\\
$^{3}$ \quad ICMC/Universidade de S\~ao Paulo, S\~ao Carlos/SP 13566-590 Brazil; moacir@icmc.usp.br\\

$^{4}$ \quad School of Computer Science and Information Technology, University College Cork, T12 YN62 Cork, Ireland}%We changed ``IT'' into ``Information Technology'', please confirm. #authors: Confirmed
\corres{Correspondence: r.minghim@cs.ucc.ie}

\abstract{Multiple-instance learning (MIL) is a paradigm of machine learning that aims to classify a set (\textit{bag}) of objects (\textit{instances}), assigning labels only to the \textit{bags}. This problem is often addressed by selecting an instance to represent each bag, transforming an MIL problem into standard supervised learning. Visualization can be a useful tool to assess learning scenarios by incorporating the users' knowledge into the classification process. Considering that multiple-instance learning is a paradigm that cannot be handled by current visualization techniques, we propose a multiscale tree-based visualization called MILTree to support MIL problems. The first level of the tree represents the bags, and the second level represents the instances belonging to each bag, allowing users to understand the MIL datasets in an intuitive way. In addition, we propose two new instance selection methods for MIL, which help users improve the model even further. Our methods can handle both binary and multiclass scenarios. In our experiments, SVM was used to build the classifiers. %confirm meaning retained %authors: meaning retained
	With support of the MILTree layout, the initial classification model was updated by changing the training set, which is composed of the prototype instances. Experimental results validate the effectiveness of our approach, showing that visual mining by MILTree can support exploring and improving models in MIL scenarios and that our instance selection methods outperform the currently available alternatives in most cases.}


\begin{document}
\section{Introduction}
\label{sec:intro}

Many machine learning problems can be solved by standard supervised learning techniques, in which an object is represented by a single feature vector~\cite{mello2018machine}. However, there are problems in which the target of the classification is a set of several instances, each one represented by a separate feature vector. This is the case of multiple-instance learning (MIL)~\cite{Fu2011MILIS}. In MIL, an object, called a \textit{bag}, contains a set of \textit{instances}. MIL was introduced in~\cite{Dietterich1997} to solve the problem of drug activity prediction, but many other studies have already applied this approach successfully, such as image classification~\cite{Amores2015MILDE}, cancer detection via images or sequences~\cite{astorino2020melanoma,XIONG2021MILReceptor}, text categorization~\cite{Ray2005SVMMIL}, speaker recognition~\cite{Reynolds00speakerverification} and web mining~\cite{Zafra2008webMIL}. Amongst the characteristics of problems that are fit to be solved by MIL approaches are those in week supervision scenarios that do not work well with standard machine learning pipelines \cite{CARBONNEAU2018Survey}.
Although multi-class classification is possible with MIL, most studies addressed only binary classification where a bag can belong either to a positive or negative class: a bag is labeled as positive if it contains at least one positive instance; otherwise, it is labeled as a negative bag.

Different supervised methods have been proposed to handle the MIL problem~\mbox{\cite{Andrews2003supportvector, 2009ZhouyuISMIL, Chen2009MedoidsMIL, Xiao2014SmileFramework}}. A widely used strategy is to convert the multiple-instance problem into a classical supervised learning problem by selecting a single feature vector (instance) among the several in each bag. This instance is often called instance prototype, which is later used to represent the bag both in training and classification steps, assuming it is sufficient to represent it correctly \cite{Yixin2006MILES, Fu2011MILIS, Huang2012SalientInst}. Take, as an example, the image classification case where each image would be considered as a bag and its segmented regions (represented by separate feature vectors) as instances. Applying this type of strategy, the hypothesis would be that one region in each image--instance prototype can represent the whole image and distinguishes that "bag'' from the others. 
Selecting a prototype by its relevance often needs to account for different factors such as similarity measures and the bias of the importance selection algorithm~\cite{ponti2019supervised}. In this context, users should play a central role in defining the selection criteria and adjusting parameters for the target model.

Visualization techniques have been successfully employed to help users in standard classification tasks \cite{Keim1996, 2007HuangVisualClass, 2012KeBingVisualClass, Paiva2015VisualClassification}. However, these techniques are not directly applicable to the case of multiple-instance data for two reasons: first, they do not scale well, which is a problem since the MIL dataset is often large due to the granularity of instances. Second, visualization methods are often designed to visualize all instances in the same space. An adequate visualization for this task should distinguish between bags and instances, reflecting the typical structure of an MIL dataset in the same layout. 

Unlike previous approaches, in this paper, we propose the use of visualization to support user intervention in the multi-instance classification pipeline. We target the tasks of instance prototype selection and training set building by the analyst, performed interactively after a preliminary solution has been obtained by an automatic procedure.
% Our target audience are the analysts who perform tasks like instance prototype selection and training set building, that are performed interactively after a preliminary solution has been obtained by an automatic procedure 
The central visualization in the approach is the MILTree, a multi-scale visualization technique based on the Neighbor-Joining similarity tree~\cite{Cuadros2007}. In the first level, only bags are placed, reducing the amount of data to be visualized compared to laying out all instances on the visual plane. The second level projects the instances belonging to a single bag, allowing the user to explore its contents. Each bag in the first level is connected to its instances in the second level, producing an intuitive structure that facilitates data analysis. 

In addition to the visual support to MIL, we also propose two new instance selection methods with multiclass support: one integrating the MILTree with Salient Instance Selection (MILTree-SI) and the other with Medoids Selection (MILTree-Med). MILTree-SI is based on the MILSIS method~\cite{Huang2012SalientInst}, which assumes that negative bags only have negative instances, while our method considers that negative bags can have positive instances as well. This adapts it for applications such as image and text. The MILTree-Med method uses the k-Medoids clustering algorithm to partition the unlabeled instances in an attempt to find positive and negative clusters, thus identifying the instance prototypes as their medoids.

% This paper addresses both binary and multiclass problems. We also propose a new methodology that allows users to take part in every step of the multiple-instance classification process and improve the training set. 

In that context, the main contributions of this work are:
\begin{itemize}
\item MILTree---a novel tree layout for multiple-instance data visualization;
\item MILTree-Med and MILTree-SI---two new methods for instance prototype selection;
\item MILSIPTree---a visual methodology to support multiple-instance data classification.
\end{itemize}

%The remainder of this paper is organized as follows. Section~\ref{sec:relatedwork} briefly reviews related works. Section~\ref{sec:precon} presents the background and related concepts about multiple-instance learning, visual data mining and visualization techniques related to our work. Section~\ref{sec:method} presents the MILTree layout as well as the MILTre-Med and MILTree-SI instance prototype selection methods. Section~\ref{sec:applications} presents MILSISTree methodology and reports three case studies of multiple-instance datasets, in order to demonstrate the usefulness of the proposed methodology. Section~\ref{sec:experiments} presents numerical experiments performed to evaluate our proposed methods in different multiple-instance datasets, as well as studies on properties of the tree. Section~\ref{sec:UsabilityTesting} presents a user study on the use of the tree. Section~\ref{sec:StatisticalAnalysis} presents a statistical analysis. Finally, Section~\ref{sec:conclusions} presents our observations regarding these novel methods. 

\section{Related Work}
\label{sec:relatedwork}

The first studies in the MIL field included the Diverse Density (DD) \cite{Dietterich1997}, DD with Expectation Maximization (EM-DD) \cite{Zhang2001EMDD} and MI-SVM \cite{Andrews2003supportvector}. Later, several methods were proposed using instance selection strategies, such as MILES \cite{Yixin2006MILES}, MILIS \cite{Fu2011MILIS} and IS-MIL \cite{2009ZhouyuISMIL}. These methods tackle the MIL problem by converting it into regular supervised learning. This is carried out by choosing instance prototypes (IP) for each bag, which then can be used to learn a classifier. MILSIS \cite{Huang2012SalientInst} is also a method based on instance selection that aims to identify instance prototypes named Salient Instances, which are \textit{true} positive instances in positive bags.

When visualization, data mining and machine learning are studied in the same context, it is possible to provide better tools for exploring and understanding data~\cite{Campello2015}. Studies relating these topics become even more important in the case of unstructured data since we need to obtain representations of those objects in reduced dimensions \cite{Ponti2016, Yu2016}.
There are several related studies aiming at supporting the classification process by means of visual tools \cite{Keim1996, 2007HuangVisualClass, 2012KeBingVisualClass, Paiva2015VisualClassification}. To process multidimensional data, previous visual mining approaches have employed both multidimensional projections and trees.

Multidimensional projections map high dimensional vectors into points in a low dimensional space, such as 2D or 3D~\cite{tejada2003}. The result of this projection is a point placement on a plane, normally corresponding closely to a similarity relationship so that an instance is likely to be placed close to other similar instances and far from those that are not similar~\cite{Ward2005}. 
A large variety of techniques can be used to perform projection, such as Principal Component Analysis (PCA)~\cite{Jolliffe2002}, Multidimensional Scaling (MDS)~\cite{Cox2001}, Least Square Projection(LSP)~\cite{Paulovich2008}, Local Affine Multidimensional Projection (LAMP)~\cite{Joia2011LAMP}, tSNE and UMAP. Although projections have improved greatly in precision and performance, they are prone to producing overlapping points, causing clutter that hampers the interaction for datasets that are reasonably large. 

On the other hand, similarity trees enforce the separability of the points by including edges between elements and causing branches between groups of similar points. They are constructed from the distances between the instances to be displayed. The Neighbor-Joining (NJ) method, originally proposed for re-constructing phylogenetic trees~\cite{Cuadros2007}, is of particular interest. NJ builds unrooted trees, aiming at minimizing the tree length and number of tree branches by finding pairs of closed instances and creating a branch from them. In this paper, we employ an improved NJ tree layout algorithm~\cite{Paiva2011Tree} that runs faster than the original NJ~\cite{Cuadros2007}.

%In \cite{2012KeBingVisualClass}, the authors present a visual approach for classification of multivariate data based in Hypothesis-Oriented Verification and Validation by Visualization (HOV$^{3}$). They employ HOV$^{3}$ to project high dimensional data to a 2D space allowing users to build a visual classifier from a training dataset based on the data projection. Then the user classifies unlabeled data points employing the visual classifier by projecting the mixed unlabeled dataset and data points of the visual classifier with the measure vector. 

%Another relevant study \cite{Paiva2015VisualClassification} provides an approach to visually support users in classification tasks such as training set selection, model creation, application and verification as well as a classifier tuning, supported by a similarity-based form of visualization. The visualization techniques used to support classification are multidimensional projections and Neighbor-Joining trees. Visualization is useful in particular to explore data such as text and images, for example by allowing to project their representations in a 2D plane showing relationships between instances and their different categories.

Our paper presents a novel contribution by extending the notion of visually supporting classification tasks to the case of multiple-instance learning, providing a methodology that assigns visual layouts to MIL tasks. We use a novel multi-level NJ tree, allowing users to explore MIL datasets, select the training set, create a model, visualize classification results, as well as update the current model using novel methods, as detailed in the next section.

\section{Background and Related Concepts}
\label{sec:precon}

In this section, we briefly present the main concepts regarding multi-instance learning and visual support for classification tasks.

\subsection{Multiple Instance Learning}
\label{sec:mil}
In supervised learning, we define classification as follows: given an instance space $\mathcal{X}$ (also called input space) composed by individual instances---each one is represented by a feature vector---and a label space $\mathcal{Y}$ (output space) composed by classes that can be assigned to each sample, the task is to build a classifier, i.e., a map $f : \mathcal{X} \rightarrow \mathcal{Y}$. The classifier is often obtained by using a training set of instances as input with their corresponding true~labels~\cite{mello2018machine}.

In MIL, an object is represented by different parts, although the object itself has its own label, each part may have, in principle, different labels. This causes the classical supervised learning definition to perform poorly in multi-instance scenarios. %The problem is thus addressed by the Multiple Instance Learning (MIL) approach, which allows assigning a single label to a bag object (set of individual instances), even though instances inside a bag can have different labels. %\cite{Amores2013}

Formally, a multiple-instance learning algorithm learns a classifier $f_{MIL} : 2^\mathcal{X} \rightarrow \left\lbrace -1, +1 \right\rbrace$, taking as input a dataset composed by bags $B_{i}$, with $i=1,\cdots, n$. Each bag contains a set of instances; the $\emph{j}^{th}$ instance inside a given bag is denoted as $B_{ij}$, and $j=1,\cdots,n_i$, in which $n_i$ is the number of instances inside the bag $B_i$. Considering a binary classification scenario, positive and negative bags are denoted, respectively, by $B_{i}^{+}$ and $B_{i}^{-}$.
For the sake of simplicity, we will denote a bag as $B$ when it represents either positive or negative bags. 

One of the crucial steps in this process is how to assign a label to the bag given the labels of its instances. The first MIL study~\cite{Dietterich1997}, related to drug activity, assigned a positive label to a bag that has at least one positive instance. This approach makes sense for that application but has not been successful for other datasets~\cite{Fu2011MILIS}. The state-of-the-art strategies often select a {\em prototype} instance in order to represent the bag, and there are many different heuristics that can be used to perform this task. 
%In this context, visualization approaches can shed some light on the formation of bags, on relevant instances, on the relationship amongst the bags in the dataset, and on how to improve the existing strategies.

In the following sections, we will give further details about two strategies for instance selection employed in multiple-instance learning. Moreover, we discuss ideas for visualization and interaction to improve prototype instance selection.

\subsection{Instance Prototype Selection}
\label{subsec:relatedConcept}

\subsubsection{Salient Instance Selection Strategy}
\label{subsec:saliente}

%In classical definition of MIL~\cite{Dietterich1997}, a positive bag contains at least one positive instance, whereas a negative bag contains negative instances only. 
In MILSIS \cite{Huang2012SalientInst}, the authors perform prototype selection in positive bags, obtaining Salient Instances via ``Rough Selection'' and ``Fine Selection''. 

{\em Rough Selection} obtains two \textit{optimal} positive instances from all positive bags, which are basically those with the highest and lowest values of salience. First, all instances in negative bags are grouped in a set $B^{-}$. Then, the salience $Sal(B^{+}_{ij})$ for each instance in a given bag $B^{+}_{i}$ is computed as follows:
% Numbered Equation
\begin{equation}
\label{equation1}
Sal(B_{ij}) = \sum_{B_{ik} \in B_{i} \setminus \left\{B_{ij}\right\} } d(B_{ij}, B_{ik}),
\end{equation}
where $d(.,.)$ is the Euclidean distance function. A high salience value indicates that the instance is different from the other instances in the bag. 
%Please check through the paper if indentation should be added to the first line the after equations.
% Algorithm
%authors: Indention is suppressed from the text right after an equation. The indention is only added when we start a new paragraph.

\begin{algorithm}[h]
\SetAlgoNoLine
\KwIn{$B$, \textit{SalNum}}
\KwOut{Salient instances (prototypes) \textit{T}.}
$B^{-} = \{B_{rt}|B_{rt} \in  B^{-}_{r}, r=1,2,...,n^{-} \}$\;
$maxDist=0$\;
    \tcp{Rough Selection}  
    \For{$i=0$ to $n^{+}$
    }{
     	Compute $Sal(B^{+}_{i1})$ for each instance in $B^{+}_{i}$ \tcp*[l]{(see Equation~\ref{equation1})}
 		Re-sort all instances in $B^{+}_{i}$ in descending order of salience.\;
 		Compute $D(B^{+}_{i1}, B^{-})$ and $D(B^{+}_{im}, B^{-})$\tcp*[l]{(see Equation~\ref{equation3})}
      \eIf{$D(B^{+}_{i1}, B^{-})$ $>$ $D(B^{+}_{im}, B^{-})$ and $D(B^{+}_{i1}, B^{-}) > 
maxDist$\;
      }{
        $maxDist = D(B^{+}_{i1}, B^{-})$ \;
		$optPosInst = B^{+}_{i1}$\;
      }
      {
      \If{$D(B^{+}_{im}, B^{-})$ $>$ $D(B^{+}_{i1}, B^{-})$ and $D(B^{+}_{im}, B^{-}) > 
maxDist$ }
	{
     	 $maxDist = D(B^{+}_{im}, B^{-})$ \;
		 $optPosInst = B^{+}_{im}$\;
      }
      
      }
     }
     \tcp{Fine Selection}  
   $optNegInst = \arg\max_{j \in B^{-}} d(j,optPosInst)$\;
\For{$i=1$ to $n^{+}$}{
	\eIf{$d(B^{+}_{i1}, optNegInst)$ $>$ $d(B^{+}_{im}, optNegInst)$
      }{
        $T =  T \cup B^{+}_{i1}, ...,B^{+}_{iSalNum}$ \tcp*[l]{Set of salient instances}
      }
      {
       $ T =  T \cup B^{+}_{i(m-SalNum+1)}, ...,B^{+}_{im}$ \;
		}

}
Return $T$\;
%\Return $T$      
\caption{MILSIS: Salience Instance Selection.}
\label{alg:euclid}
\end{algorithm}

After computing saliences inside $B^{+}_{i}$, instances are sorted from the maximum ($j=1$) to the minimum ($j=m$) salience values. This is used to estimate the probability that $B^{+}_{i1}$ (maximum) and $B^{+}_{im}$ (minimum) instances are positives given the set $B^{-}$ (see Equation~\ref{equation2}), and then select an \textit{optimal} positive instance, which will represent $B_{i}$:

\begin{equation}
\label{equation2}
 Pr(l(B_{ij}) = +1 | B^{-}) = 1- \exp(-D(B_{ij},B^{-}) / {\sigma}^{2} ),
\end{equation}
where $l(.)$ is the label function, $\sigma$ is a scaling factor larger than 0; $D(B_{ij},B^{-})$ is the minimum distance between $B_{ij}$ and all instances in $B^{-}$:
\begin{equation}
\label{equation3}
 D(B_{ij},B^{-}) = \min_{B_{rt}\in B^{-}} d(B_{ij}, B_{rt}),
\end{equation}

Since $Pr(l(B_{ij}) = +1|B^{-})$ is proportional to $D(B_{ij},B^{-})$~\cite{Huang2012SalientInst}, from Equation~\ref{equation3}, it is possible to estimate how likely an instance is to be labeled as positive or negative by its distance to the set of negative instances. Finally, the probabilities of each bag are compared in order to find the \textit{optimal} positive instance, i.e., the one with maximum distance to $B^{-}$.

In {\em Fine Selection}, an \textit{optimal} negative instance is selected from $B^{-}$ by its maximum distance to the optimal positive ones. Then, starting with the optimal positive instances obtained in the rough selection: $B^{+}_{i1}$ and $B^{+}_{im}$, it finds a \textit{true} positive instance in each bag. These \textit{true} positive instances will be part of the Salient Instances. Algorithm~\ref{alg:euclid} summarizes the whole procedure. Note that $B = B^{+} \cup B^{-}$; $n^+$ and $n^-$ are, respectively, the number of instances inside positive and negative bags; and \textit{SalNum} is the number of salient instances. 
  
\subsubsection{Medoids Instance Selection Strategy}
\label{subsec:medoids}
The Medoids Instance Selection \cite{Zhang2009medoidsMIL, Zhou2007medoidsMIMLL} strategy computes the medoid for each bag using the $k$-Medoids algorithm. Each medoid will represent a bag, and then the multiple-instance learning task is reduced to a traditional supervised learning task. The k-Medoids algorithm is adapted to cluster the multiple-instance data so as to partition the unlabeled training bags into $k$ groups. After that, it re-represents each bag by a $k$-dimensional feature vector, where the value of the $i^{th}$ feature is the distance between the bag and the medoids of the $i^{th}$ group. 
In other words, the medoid of each bag has the minimum average distance to the other bags in the same group, with all bags represented as $k$-dimensional feature vectors in a regular supervised learning approach.

% \subsection{Visual Data Mining and Information Visualization}
% \label{sec:visualdatamining}

% Visual data mining aims at incorporating the perception of humans in data mining processes, combining user knowledge with computational capabilities of algorithms~\cite{Rashid2012}, either in chosen stages of the processes or throughout the data mining pipeline~\cite{Paiva2015VisualClassification}. Common stages include: initial data exploration, model building, model testing, model applying, and model deployment. 

% % Visualization is useful in particular to improve data exploration on multidimensional datasets---which is often the case when dealing with complex data such as text and images~\cite{Joia2011LAMP}---because it encodes proximity, enclosure, similarity, connection, and continuity between data in a visualization, allowing the user to get insight into data and draw conclusions. 
 
\section{Visual Multiple-Instance Learning}
\label{sec:method}
Our approach to tackle the multiple-instance learning problem consists of two main features: a tree-based visualization to encode the MIL data (including instances and bags representations), coupled with new heuristics  based on that visualization, to convert MIL into a standard machine learning problem.

\textls[-10]{The data under analysis can be visualized in the \textit{bag space} or the \textit{instance space} using MILTree. We also identify prototypes for each bag, which allows training a classifier using those prototypes.~Two~methods were designed to identify prototypes: MILTree-SI and MILTree-Med, both using the MILTree visualization proposed in this work. }
In this section, after defining additional notation, we describe the MILTree layout and then the MILTree-SI and MILTree-Med instance prototype selection methods.  
 
\subsection{Additional Notation}
\label{subsec:notation}
In addition to the notation described in Section~\ref{sec:mil}, for each bag we designate two special instance prototypes, denoted by $B_{protoProj}$ and $B_{protoClass}$. The $B_{protoProj}$ is used for visualization purposes, denoting the instance prototype that will be used to map bags in the MILTree's bag space layout, while $B_{protoClass}$ denotes the prototype used to create the classification model. Initially they are the same; however in order to keep the same visualization layout while updating the classification model, $B_{protoClass}$ can change, but $B_{protoProj}$ does not change to preserve the MILTree layout throughout the visual mining~process.

\subsection{Creating a Multiple-Instance Tree (MILTree)}
\label{subsec:miltree}

The improved NJ algorithm~\cite{Paiva2011Tree} begins with a star tree formed by all $m$ objects on the distance matrix, represented by leaf nodes arranged in a circular configuration and connected by branches to a single central node. Then, it iteratively finds the closest neighboring pair among all possible pairs of nodes by the criterion of minimum evolution, which attempts to minimize the sum of all edge lengths for all nodes of the tree. 
Afterwards, the closest pair is clustered into a new internal node, and the distances from this node to the remaining ones are computed to be used in subsequent iterations. The algorithm stops when $m-2$ virtual nodes have been inserted into the tree, i.e., when the star tree is completely resolved into a binary tree. 

Algorithm~\ref{alg:alg1} illustrates the NJ tree procedure, which starts by computing the depth of the divergence for each node, i.e., the sum of the distances from instance $i$ to all other~nodes:
\begin{equation}
\label{paresNos}
r_{i} =  \sum_{j \not= i}D_{i,j}.
\end{equation}
Then, it computes a new distance matrix based on the divergence $r_{i}$ in order to find the closest pair of nodes $i,j$:
\begin{equation}
\label{reducaoMatriz}
c_{i,j} = D_{i,j}-\frac{r_{i} + r_{j}}{-2},
\end{equation}
After finding the pair of nodes $i,j$, a virtual node $u$ is created as a parent of both $i$ and $j$. The length of the edge connecting $u$ to $i$ is:
\begin{equation}
\label{tamanhoRamos1}
s_{i,u} = \frac{D_{i,j}}{2}+\frac{r_{i} + r_{j}}{2(m-2)},
\end{equation}
and the length of the edge connecting $u$ to $j$ is:
\begin{equation}
\label{tamanhoRamos2}
s_{j,u} = D_{i,j}-s_{i,u}.
\end{equation}
Finally, the pair of nodes $i,j$ is replaced by $u$ in the matrix $D$, and the distances between $u$ and all others nodes are computed by Equation (\ref{calculodistanciaNJ}), where $k\not=i$, $k\not=j$ and $k=1..m$:
\begin{equation}
\label{calculodistanciaNJ}
D_{k,u} = \frac{D_{i,k}+D_{j,k}-D_{i,j}}{2}
\end{equation}
The algorithm then iterates, finding and joining pairs of nodes until $m-2$ virtual nodes are inserted into the tree.%Please check through the paper if indentation should be added to the first line the after equations.
%Authors: Indention is suppressed from the text right after an equation. The indention is only added when we start a new paragraph.

% Algorithm
\begin{algorithm}[h]
\SetAlgoNoLine
\KwIn{Similarity matrix $D$.}
\KwOut{Phylogenetic Neighbor-Joining Tree.}
$v' = m$ \tcp*[l]{Where $m=D.size$}
\While{$v'>2$
    }{
      Compute $r_i$ for $i=1..m$, and find  $c_{i,j}$ which is the closest pair of instances $i,j$ \tcp*[l]{(see Equation (\ref{paresNos}) and (\ref{reducaoMatriz})).}
		 Create a new virtual node $u$\;
		 Compute the lengths of edges $(u,i)$ and $(u,j)$ \tcp*[l]{(see Equations (\ref{tamanhoRamos1}) and (\ref{tamanhoRamos2})).}
         
		 Replace $i,j$ by $u$ and update the distance  matrix $D$ with the new node $u$ \tcp*[l]{(see Equation (\ref{calculodistanciaNJ})).}
 		Define $u$ as parent of both $i$ and $j$\;
		 $v' = v'-1$\;
    }
      %Return  Tree\;
\caption{NJ tree Algorithm.}
\label{alg:alg1}
\end{algorithm}

To map multiple-instance data to a visual tree structure, the MILTree was developed as a two-level NJ tree, with bags and instances projected in different levels. We group the instance matrix data by the previously known bags using the instance prototypes as nodes of the second level of the NJ. The layout of the tree projects the data into visual space.

\begin{figure}[H]
\includegraphics[width=0.55\textwidth]{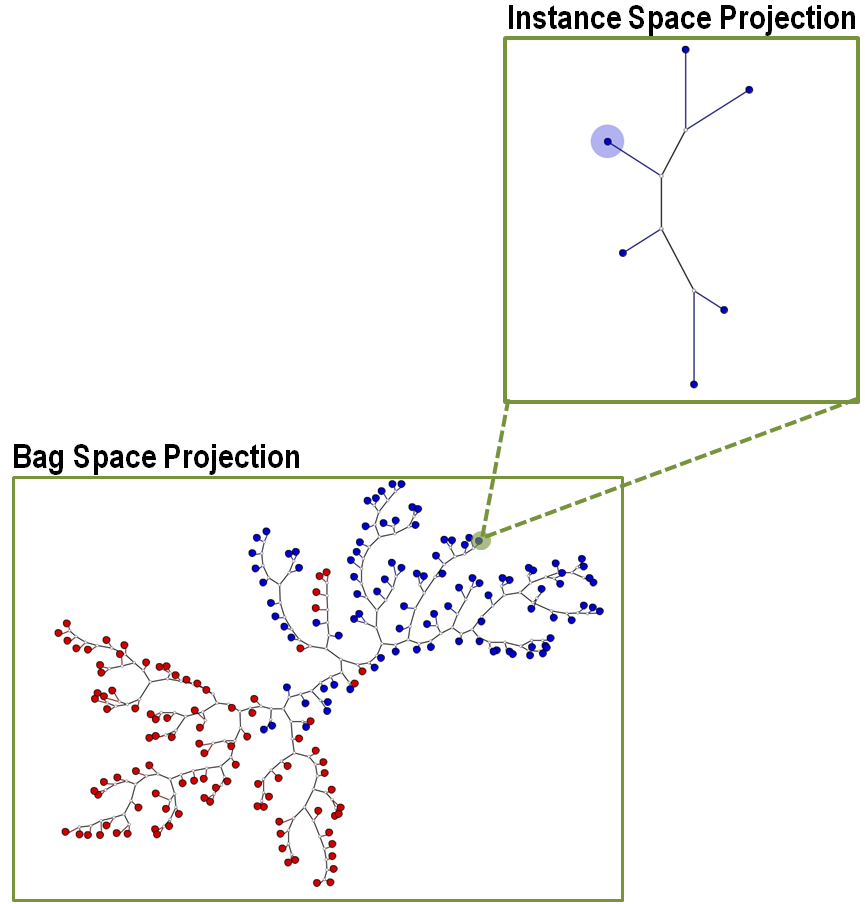}
\caption{Layouts of Bag and Instance Spaces for the Corel-1000 subset (with instance prototypes highlighted) in the MILTree, with a total of 200 bags and 824 instances.}
\label{spacesProjection_MILTree}%MDPI: Please note that figures and tables should be insert AFTER where it was first mentioned. The size of subfigures will be adjusted appropriately to ensuring the clarity and readability of the image. Changes to the position of figures and tables may occur during the final steps.
%Authors: We make sure that figures and tables are inserted after where it was first mentioned.
\end{figure}

A subset of the Corel-1000 dataset is used to illustrate the bag and instance projection levels in Figure~\ref{spacesProjection_MILTree}. In this dataset, each image is a bag, composed of feature vectors (instances) that are extracted from disjoint regions of the image, with an average of 4.5 instances per bag. In the first-level (bag space projection) the red points represent positive bags---100 images of the flower category---and the blue points represent negative bags---100 images selected uniformly from the remaining categories of the dataset. When a bag is selected, its instances are projected in the second level of the multiple-instance tree (instance space~projection). 

\textls[-10]{Algorithm \ref{alg:algMILTree} contains the complete procedure to build the MILTree, which starts by grouping the matrix data $D$ in bags $B_{i}$ (first {\tt for} loop), where $i$ denotes the index of bags. We iterate over the instances in the matrix $D$, each line $D_{m}$ representing the distance from instance $m$ to all other instances. 
In each iteration, a new bag $B_{i}$ is created, and all its instances $B_{ij}$ are added to $B_{i}$, where $j$ denote the index of instances belonging to some bag $B_{i}$. }

Afterwards, we compute $B_{i,protoProj}$ and $B_{i, protoClass}$ for each $B_{i}$ using either MILTree-SI or MILTree-Med (second {\tt for} loop). Remember that $B_{i, protoProj}$ is used in MILTree and $B_{i,protoClass}$ in the classification process. Every $B_{i, protoProj}$ is included in a set $P$, which will be used later to create a bag distance matrix in the NJ tree procedure. Finally, $B_i \forall i$ are processed by the MILTree using $P$, creating the bag space projection.

\IncMargin{1em}
\begin{algorithm}[h]
\SetAlgoLined
\SetKwFunction{NJTree}{NJTree}
\KwIn{Similarity matrix $D$.}
\KwOut{Multiple-Instance Tree.}

\tcp{Creating bags:}  
  \For{$m=0$, $i=0$ to $m<D.size$
    }{    
    \For{$j=0$ to $B_{i}.size \in D$
    }{
   		$B_{ij}$ = $D_{m}$.\;
		Add $B_{ij}$ to $B_{i}$. Where $B_{ij} \in D_{n}$ , $i=idBag$, $j=idInstances$.\; % of a bag.
 		Connect $B_{i}$ as parent of instance $B_{ij}$\;
 		$m=m+1$ \tcp*[l]{Taking the next instance}
    }    %\State $i=i+1;$ //create a new bag.
    }
    
\tcp{Computing instance prototypes:}  
    \For{$i=0$ to $B.size$
    }{
		 \tcp{MILTree-Med/MILTree-SI prototypes selection methods (see Section~\ref{sec:visualMILselection})}
   		 Compute $B_{i, protoProj}$ and $B_{i, protoClass}$ using MILTree-Med or MILTree-SI.\;
 		$P = P \cup B_{i, protoProj}$ \tcp*[l]{set of all projection prototypes $B_{i, protoProj}$}
    }
    
\tcp{Projecting bags in the bags space projection of MILTree:}  
		$B.tree = $\NJTree{$B_i$,$P$}, $i =0,1,2,....,B_{i}.size$\; 
 		%Compute the promotion the NJTree $B.tree$\; 
		%\For{each $B_{i}$ in $B$}
		%\State $B_{i}.tree = NJ(B_{i},B_{i, protoProj});$
		%\State Compute the promotion of the NJTree $B_{i}.tree$;
		%\EndFor\\
		Return $B.tree$.\; 
		%\State \textbf{return} $T$\Comment{The gcd is b}
		%\EndProcedure
		%\Return $T$   

\caption{MILTree Algorithm.}
\label{alg:algMILTree}
\end{algorithm}
\DecMargin{1em}

\textbf{Instance projection:} Since our MILTree has two levels, when a user interacts with a bag $i$, the instances $B_{ij}$ will form an instance space projection $B_{i}.tree$. To create $B_{i}.tree$, the NJ-Tree algorithm takes as input the instances of $B_{i}$ and the prototype $B_{i, protoProj}$.

%\textbf{Promotion NJ Tree:} the NJ tree is formed by $m$ objects and $m-2$ virtual nodes. In large datasets, the space projection can be overloaded due to the high quantity of virtual nodes. In order to alleviate this issue,~\cite{Paiva2011Tree} presents a fast pattern-replacement algorithm called promotion that replaces an internal node with a leaf whenever a certain configuration of nodes exists. In our work, the decision of promoting the tree depends on the user. It is advisable to apply the promotion only when the visualization is overloaded.

\subsection{Instance Prototype Selection Methods}
\label{sec:visualMILselection}

We propose two new prototype selection methods, MILTree-SI and MILTree-Med, based on those proposed by \cite{Huang2012SalientInst}, as described in Section~\ref{subsec:relatedConcept}. Both the SI and Med approaches compute two prototypes per bag: $B_{ix}$ and $B_{iy}$. The first, $B_{ix}$, is used both in the visualization and to build the classification model, while $B_{iy}$ is an alternative prototype that can be used to update the classification model. Computing $B_{iy}$ offers an option to automatically change the bags' prototypes that are poorly represented by $B_{ix}$, for example, those that are misclassified in the training set, improving the multiple-instance classification model.

In contrast with the original method~\cite{Huang2012SalientInst}, we assume that not only positive bags but also negative bags could have both positive and negative instances. This is true for more complex data such as images and text. Consider, for example, the problem of discriminating between photos of flowers (positive) and photos of the classes person and animals (negative) in which each image is a bag and its instances are disjoint regions of the image. If we use the classic definition of a positive bag, all those images (bag) containing a flower in at least one region (instance) are considered positive; to be considered negative, the image must not contain any flower. However, in this application, we are often interested in the main object in the scene. A photo whose main subject is an animal may contain a region with a flower, for example, in the background or a person may carry a flower in a photo, although the person itself is the main object. Similar examples apply to contexts such as text, video and speech classification. Therefore, we compute both optimal positive and negative prototypes.

\subsubsection{MILTree-SI} 

\textls[-15]{The optimal negative instance is defined as the one most distant from all the true positive instances of $B^{+}$ obtained by the original Salient Instance Selection Method: $Sal(B_{ij}^-)$ is computed using Equation (\ref{equation1}), then a true negative instance is obtained for each negative bag according to Equation (\ref{equation3}). Equations (\ref{equation1}) and (\ref{equation3}) are reproduced again below for~clarity:}
\[% Unnumbered Equation
 Sal(B_{ij}) = \sum_{B_{ik} \in B_{i} \setminus \left\{B_{ij}\right\} } d(B_{ij}, B_{ik})
\]

\[
 D(B_{ij},B^{-}) = \min_{B_{rt}\in B^{-}} d(B_{ij}, B_{rt}).
\]

From all true negative instances, the optimal negative instance will be the one furthest from the set $B^{+}$. Afterwards, we select the optimal positive instance from $B^{+}$ using a similar procedure, but this time selecting the instance in $B^{+}$ with the maximum distance to the optimal negative instance found previously.

\begin{figure}[H]
\centering
\includegraphics[width=0.55\textwidth]{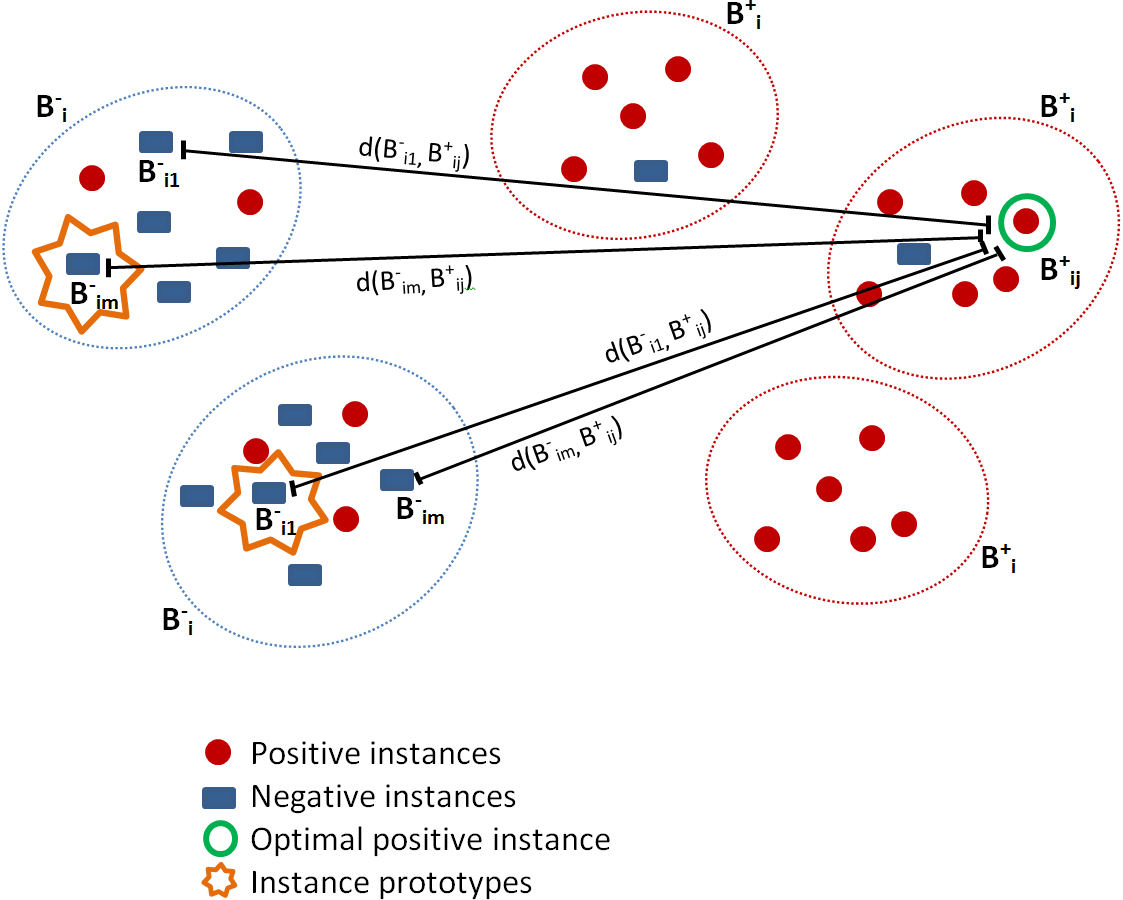}
\caption{Selection of instance prototypes on negative bags using MILTre-SI. $B^{+}_{i}$ represents positive bags, $B^{-}_{i}$ represents negative bags and $d(B^{+}_{i1}, B^{-}_{ij})$ represents the Euclidean distance between the optimal positive and a given negative instance.}
\label{MILTreeSI_representation}
\end{figure}

Finally, as we already have the optimal instances (positive and negative), we compute the instance prototypes $B_{ix}$ and  $B_{iy}$ for positive and negatives bags: $B_{ix}$ is the instance with the highest salience, and $B_{iy}$ is the next one with major salience. Figure~\ref{MILTreeSI_representation} shows the selection of prototypes (which could be $B_{ix}$ or $B_{iy}$) from negative bags using MILTre-SI.

\subsubsection{MILTree-Med} Clustering algorithms have been frequently used for selecting prototypes in a feature space defined by the instances in MIL, as presented in Section~\ref{subsec:medoids}. The original methods create a new artificial instance that represents a bag by choosing, for instance, a cluster centroid. MILTree-Med, unlike other methods, works in the instance space of each bag, selecting an actual instance without creating new ones, which better complies with the visualization scalability. Each bag is considered a cluster, and the prototype is the medoid of the cluster. Since we want to find two prototypes, the $k$-Medoids algorithm is applied with $k=2$. Since all bags may contain positive and negative instances, we want to identify potentially positive and negative clusters.

\begin{figure}[h]
\centering
\includegraphics[width=0.55\textwidth]{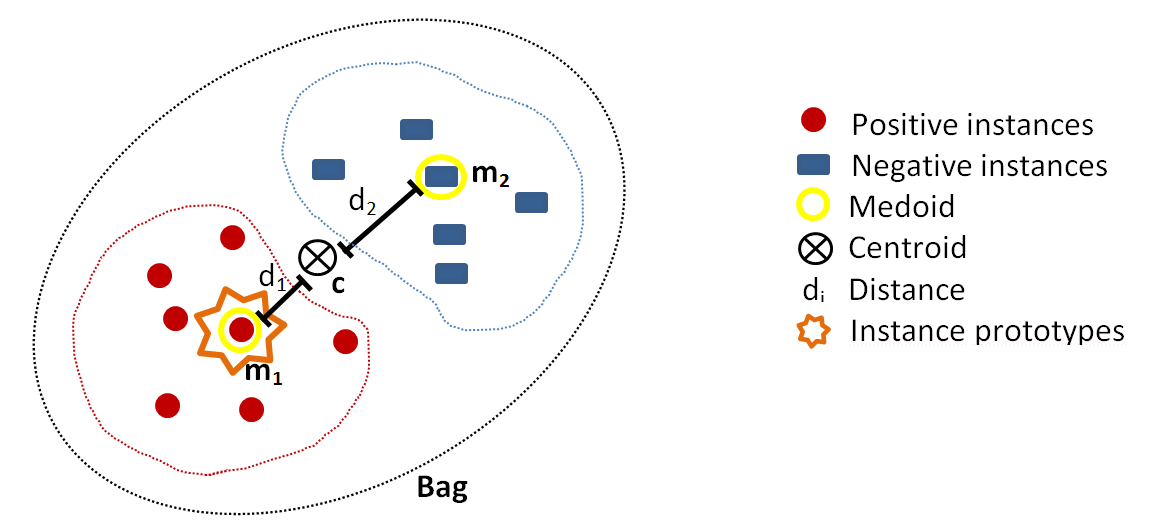}
\caption{Selection of instance prototype using MILTree-Med.}
\label{MILTree_Med_representation}
\end{figure}

The medoids of the sub-clusters are the instance prototypes $B_{ix}$ and  $B_{iy}$. Figure~\ref{MILTree_Med_representation} shows the selection of prototypes using MILTree-Med. $Bag$ represents either a positive or negative bag, $c$ is the centroid of the bag, $m_{1}$ and $m_{2}$ are the medoids of the two sub-clusters and $d$ are distances between each medoid and the centroid. 

\subsubsection{Updating Instance Prototypes Using MILTree}
To create the first MILTree layout, as well as the classifier, we set $B_{i,protoClass} = B_{i,protoProj} =B_{i,x}$, i.e., the first instance prototype selected by either MILTree-SI or MILTree-Med heuristics. The MILTree was developed so that the user can spot those bags that are poorly represented by the first prototype selection. For those bags, users can then set $B_{i,protoClass} =B_{iy}$ or manually inspect the bags to select a more representative one.

\begin{figure}[H]
\centering
\subfloat[]{\includegraphics[width=0.32\linewidth]{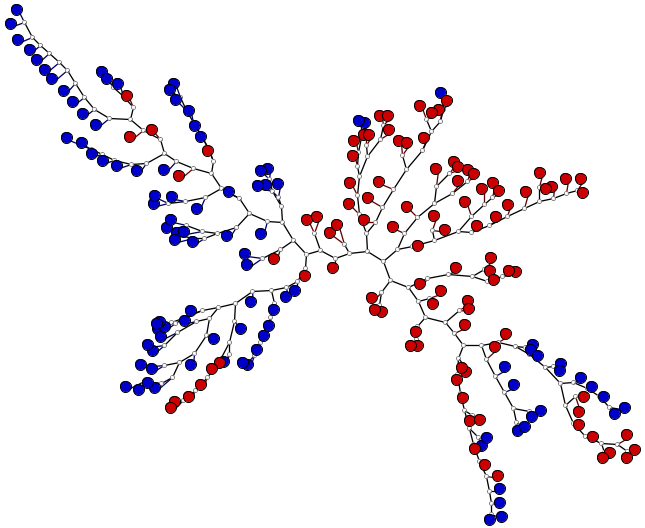}%
\label{horse_dataset}}
\hfil
\hspace{0.1mm}
\subfloat[]{\includegraphics[width=0.32\linewidth]{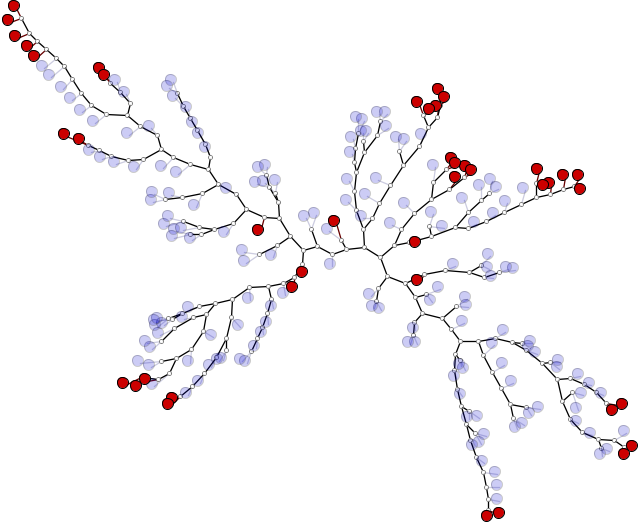}%
\label{horse_train}}
\hspace{0.1mm}
\subfloat[]{\includegraphics[width=0.32\linewidth]{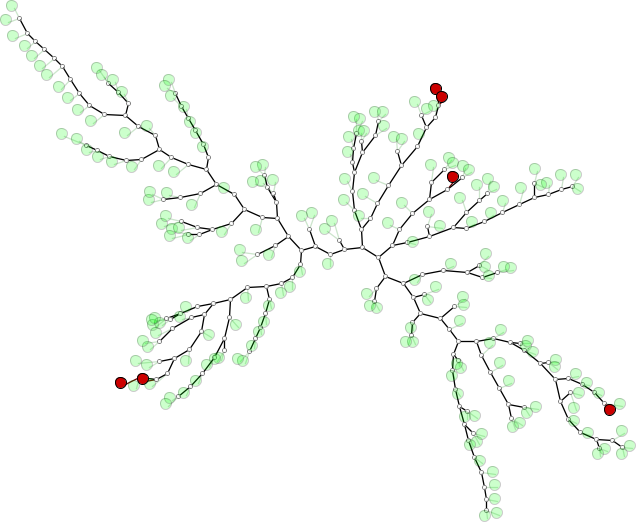}%
\label{horse_instProto}}
\caption{MILTree's bag space  layout for a subset of the Corel-1000 dataset (100 images of the horse category and 100 random images selected from the remaining categories), with the projection of its ground truth (a), selected training set (b) and \textit{InstancePrototypes ClassMatch} tree (c).}
\label{horse_data}
\end{figure}

Two visual representations are available to the user:
\begin{enumerate}
\item \textit{Prototype highlighting}: MILTree highlights the current prototype $B_{ix}$ with a darker color and also $B_{iy}$, which is the alternative prototype, with a lighter shade. Thus, by inspecting both, the user can validate $B_{protoClass}$ or update it according to their knowledge by selecting $B_{iy}$ or even another instance in the instance space layout. Figure \ref{instanceSpaceProjection} shows the instance prototypes $B_{ix}$ and $B_{iy}$ projected in the MILTree's instance space  layout. In Figure~\ref{instanceSpaceProjection}a, the SI selection method is used, and in Figure~\ref{instanceSpaceProjection}b, the medoids selection is used instead.

\begin{figure}[H]
\centering
\subfloat[Salient instance selection method.]{\includegraphics[width=1.2in,height=1.7in]{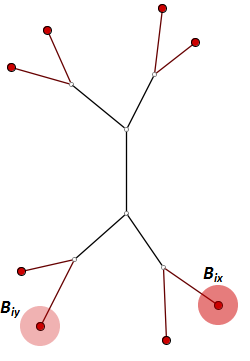}%
\label{fig_first_case}}
\hfil
\subfloat[Medoids instance selection method.]{\includegraphics[width=1.2in,height=1.8in]{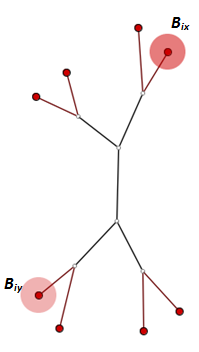}%
\label{fig_second_case}}
\caption{Methods for selecting instance prototypes $B_{ix}$ and $B_{iy}$. Both (a) and (b) project the same instances from a positive bag $B^{+}_{i}$ of the MUSK1 dataset on the MILTree's instance space layout.}
\label{instanceSpaceProjection}
\end{figure}\vspace{-6pt}

\item \textit{SVM class match tree}: the \textit{InstancePrototypes ClassMatch} tree uses color to contrast the bags that were misclassified, considering a training or validation set for which the labels are known. A similar approach has been successfully used in \cite{Paiva2015VisualClassification} and \cite{Paiva2011Tree}. In this approach, the instances $B_{protoClass}$ are used to build an SVM classifier, and the MILTree plots a layout we called  \textit{InstancePrototypes ClassMatch} tree, with colors according to the classification result: pale green for correctly classified and red for misclassified bags. Figure~\ref{horse_dataset} displays the MILTree generated for a subset of the Corel-1000 dataset, where red bags represent positive bags (images of horses) and blue bags represent negative bags (random images from other categories). To find the \textit{InstancePrototypes ClassMatch} tree for this dataset, we allow users to select a training set to create an SVM classifier. Figure~\ref{horse_train} shows the training set that was used to create the classifier; dark red bags are the ones used for training, while the pale blue ones are used as validation/test. Finally, the \textit{InstancePrototypes ClassMatch} tree shows the classification results (see Figure \ref{horse_instProto}), where dark red points are misclassified bags, probably with non-representative prototypes. Updating the prototypes will improve the model, as indicated by the results presented later in Section \ref{sec:experiments}. 
\end{enumerate}

\section{Application of MILTree to Multiple-Instance Learning Scenarios}
\label{sec:applications}
% The MILTree and the prototype selection methods SI and Med allow the user to analyze a MIL scenario by visualizing the similarity among the bags, each one represented by a prototype instance. With use of the \emph{InstancePrototypes ClassMatch} Tree, it is possible to see the errors inside the training set, change the instances that represent misclassified bags by using the alternative prototypes computed using the selection methods. The user can also manually access the contents of each bag by visualizing the instance space---in this visualization the MILTree automatically shows alternative prototypes that are automatically computed, but the user can also manually select another instance. Furthermore, by using MILTree, it is possible to visualize a complete classification process, from the creation and tuning of the training set, the classification of a test set, and also include new bags to the training set that can contribute to the model, which is called MILSIPTree methodology. 

In this section, we present three case studies that illustrate the practical usefulness of MILTree. The case studies were carried out on a Dell workstation Z620 equipped with an Intel Core CPU (E5-2690, 3.40GHz) and 16GB memory.

The first presents a binary multiple-instance image classification problem using the Corel People dataset. The second one describes a multi-class scenario using images from five classes of the Corel dataset. Lastly, the MIL benchmark dataset Musk1 is used in the third case study. More information about each dataset can be found in Section~\ref{sec:experiments}. %Each case study poses different challenges in the context of MIL as well as different visualization-based strategies to update the models and improve their performances.

\subsection{Case 1: Instance Space Layout in a Binary Classification Problem}
\label{sec:MILSIPTreeInstSpaceProj}
\begin{figure}[H]
\centering
\subfloat[]{\includegraphics[width=0.32\linewidth]{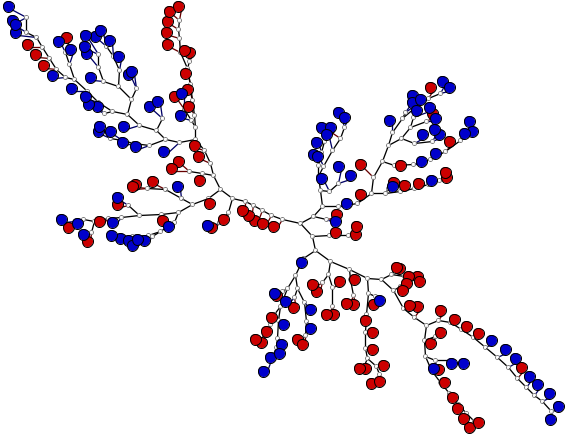}%
\label{people_dataset}}
\hfil
\hspace{0.1mm}
\subfloat[]{\includegraphics[width=0.32\linewidth]{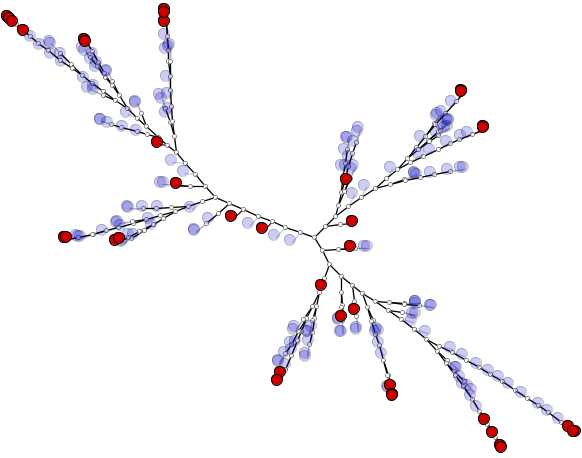}%
\label{people_train}}
\hspace{0.1mm}
\subfloat[]{\includegraphics[width=0.32\linewidth]{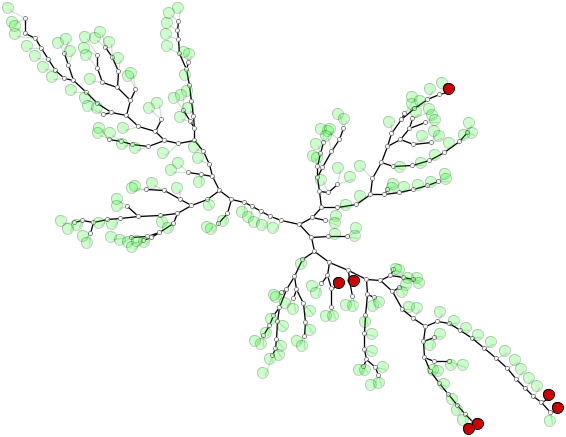}%
\label{people_instProto}}
\caption{MILTree's Bag Space Layout for the People Category of the Corel-1000 dataset, with the projection of its Ground truth (a), selected training set (b) and \textit{InstancePrototypes ClassMatch tree} (c).}
\label{people_data}
\end{figure}

We demonstrate that
an appropriate selection of instance prototypes can influence the accuracy of the classification. We use MILTree for the layout and MILTree-Med as instance prototype selection method, and the updates of prototypes are performed in the Instance Space Layout. For this case, we use the Corel People binary dataset with 200 bags (images) and 938 instances (feature vectors extracted from image regions), which includes 100 images from the class People (positive) and 100 images randomly selected from all other classes (negative) of the Corel-1000 dataset. Figure~\ref{people_data}a shows the projection of the bags using MILTree: red bags represent positive bags (images of people) and blue bags represent negative bags (images from other categories). 
\vspace{-6pt}

The first step in the classification process is to select the training set: 20\% of the images are selected to train the model, while the remaining $80\%$ is used for validation and test. Due to the nature of the NJ algorithm, MILTree positions the bags that better characterize the class they belong to as far as possible from the core of the tree (external points), while the bags located in the core of the tree (internal points) have features that overlap with other classes. This characteristic can be extremely useful in identifying a representative sample by selecting both external and internal bags to create our training set so as to build a classifier that is neither too restrictive nor too general. 
This visual selection strategy has been demonstrated to produce better results in standard supervised learning when compared with random selection strategies~\cite{Paiva2015VisualClassification}. Note that, in our representation, when the user selects a bag she/he is actually selecting its instance prototype $B_{protoClass}$. Figure \ref{people_data}b shows the selected training set for the People dataset, where red bags represent the training set and blue bags represent the test dataset. 

The selected training set is then used to create an initial classifier (in our experiments, an SVM classifier). After applying the model over the validation set, we display the \textit{InstancePrototypes ClassMatch} tree that highlights, in a contrasting color, the misclassified \textit{training} bags, whose instance prototypes are likely to be non-representative. Figure \ref{people_data}c shows the \textit{InstancePrototypes ClassMatch} tree for the People dataset, highlighting in red the misclassified points belonging to the training set. We then use the second level projection of MILTree to explore the instance space projection of those bags, aiming to improve the classification model.

To improve the MIL classification model, users have two options.  
The first option is to automatically change the $B_{protoClass}$ of all misclassified training bags identified in the \textit{InstancePrototypes ClassMatch} tree by replacing it with the alternative instance prototypes $B_{iy}$. The second option, which we show in this case study, is to visually explore the instance space projection of all misclassified bags highlighted in the \textit{InstancePrototypes ClassMatch} tree and individually choose how to update the instance prototypes.

\begin{figure}[H]
\centering
\includegraphics[width=0.8\linewidth]{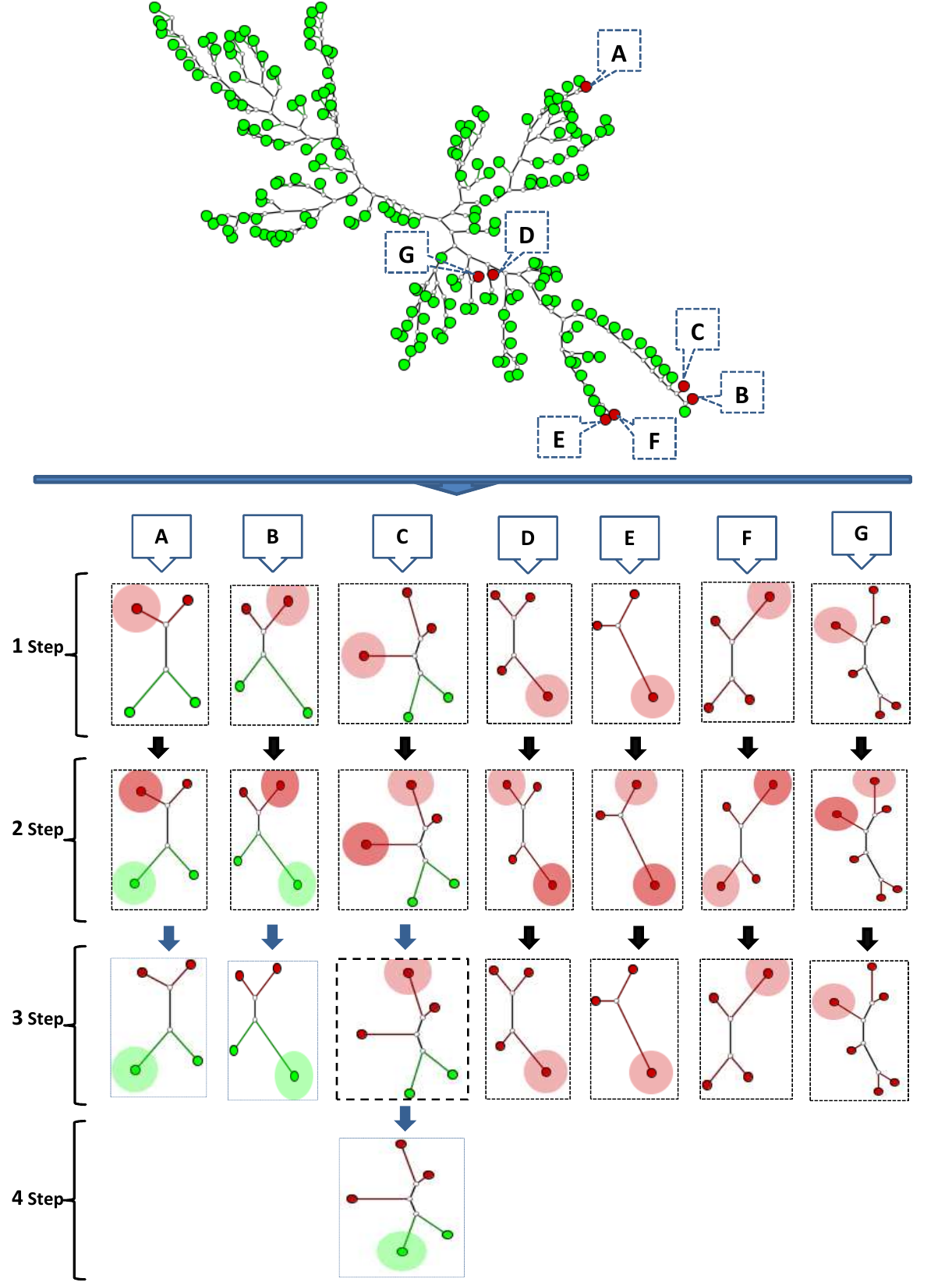}
\caption{Instance Space Layout of each bag with an unsuitable instance prototype. A, B, C, D, E, F and G represent red bags. }
\label{people_updateInstProto}
\end{figure}

Figure~\ref{people_updateInstProto} shows the instance space layout of each highlighted bag, here referred to as A, B, C, D, E, F and G. Each instance space layout is a new NJ tree formed by instances that belong to the explored bag. 
We show the four possible visual steps required to update the prototypes. Note that the first, second and third steps could be executed automatically, but by exploring it manually, the user is allowed to see and control those steps individually. For instance, the bag C required a manual selection of the correct prototype, so a fourth step is added. This allows users to explore and analyze the instances in the instance space~layout.

%Is the bold necessary? Please confirm.
%Authors: We used the italics tag instead of the bold tag to emphasize the four steps within the paragraph.
Using the example shown in Figure~\ref{people_updateInstProto}, we follow the four steps. The \textit{first step} presents the initial status: the green and red points represent, respectively, the correct classified and misclassified instances in the training set; the current instance prototypes ($B_{ix}$) are highlighted with larger circles. In the \textit{second step}, the alternative prototypes $B_{iy}$ selected by MILTree-Med are shown. In our example, it makes no sense to update the prototypes of D, E, F and G because all instances were classified in the same class. However, inside bags A, B and C, two classes are available, so users can then accept the automatic MILTree-Med update by changing the prototype to the one with the correct classification result. In the \textit{third step}, the new instance prototypes of bags A, B and C are shown, but the alternative prototype $B_{iy}$ is still misclassified in C. Due to this, a \textit{fourth step} can be carried out to manually choose a new instance prototype---in this case, a correctly classified one (green point) that is near the previous prototype.

\begin{figure}[H]
\centering
\subfloat[Classification result.]{\includegraphics[width=0.35\linewidth]{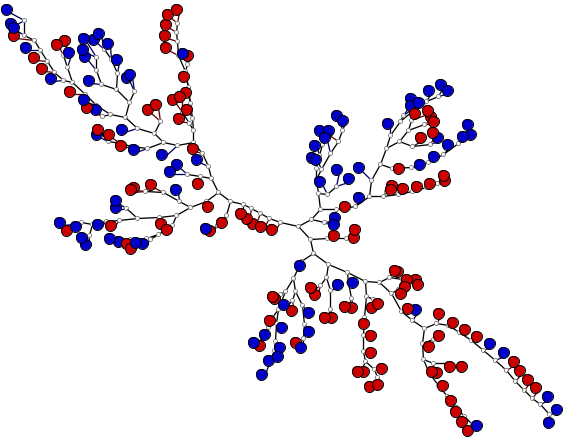}%
\label{people_classified}}
\hspace{0.5mm}
\subfloat[\emph{ClassMatch} tree.]{\includegraphics[width=0.35\linewidth]{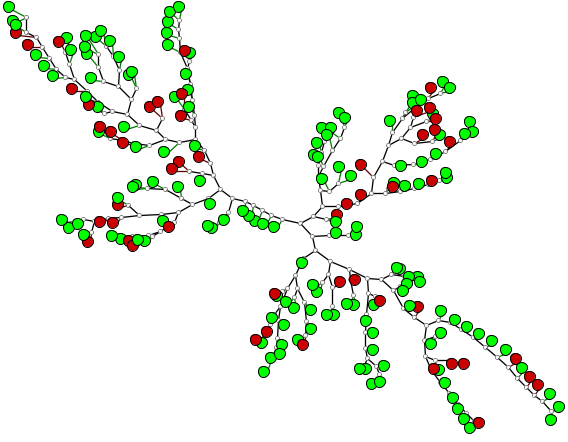}%
\label{people_classmatch}}
\caption{Classification result in the Bag Space Layout of MILTree for the People Category of the Corel-1000 dataset using a classification model with new instance prototypes (\textbf{a}) and corresponding \emph{classMatch} tree (\textbf{b}).}
\label{people_classifiedData}
\end{figure}

After updating the bags' prototypes detected through the \textit{InstancePrototypes ClassMatch} tree, we retrain the classification model. An accuracy of 72\% was achieved before the update, and by updating only three bags, it was possible to increase it to 75\%. In \mbox{Figure~\ref{people_classifiedData}a}, we show the classification results using MILTree, and its correspondent \textit{ClassMatch} tree is shown  in Figure~\ref{people_classifiedData}b. Note that the \textit{ClassMatch} tree shows the bags that were misclassified in the whole dataset, including training and validation/test sets, while the \textit{InstancePrototypes ClassMatch} tree only shows the bags that were misclassified in the training set. This case study demonstrates the positive impact of selecting representative instances on the accuracy of the classifier and that the visual exploration can play an important role by making this task easier.

\vspace{-6pt}

\subsection{Case 2: Bag Space Layout and a Multiclass Classification Problem}
\label{subsec:casetwo}
\begin{figure}[H]
\centering
\subfloat[Ground truth.]{\includegraphics[width=0.32\linewidth]{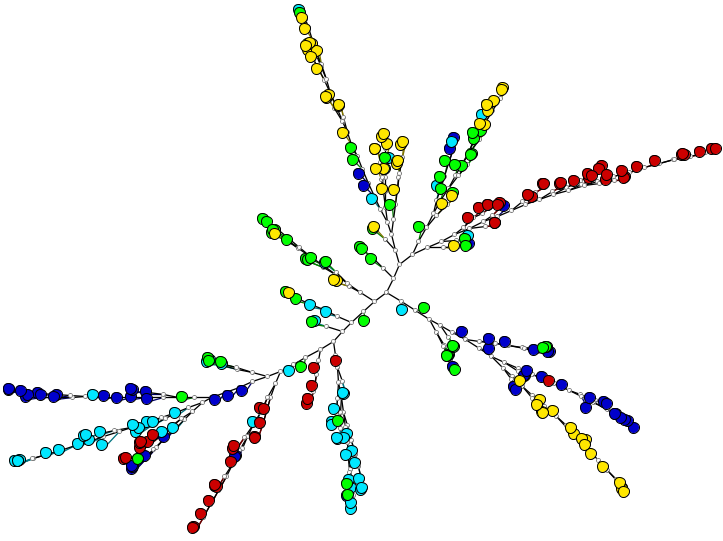}%
\label{fig_corel300_ground}}
\hfil
\hspace{0.1mm}
\subfloat[Selected training set.]{\includegraphics[width=0.32\linewidth]{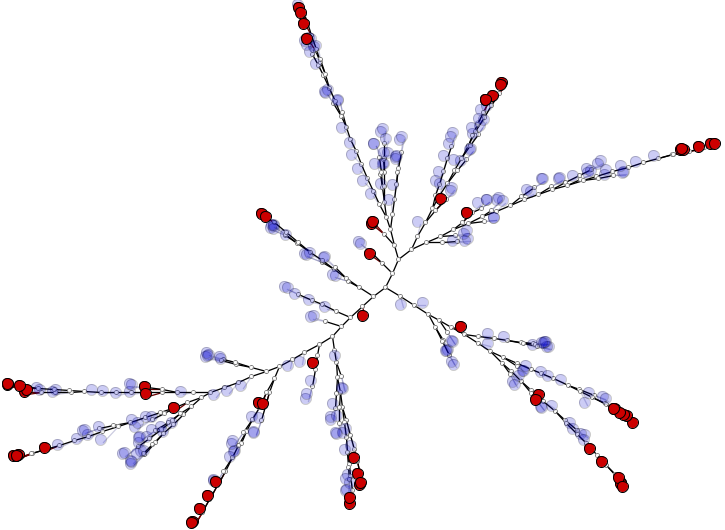}%
\label{fig_corel300_training}}
\hspace{0.1mm}
\subfloat[\textit{InstancePrototypes ClassMatch} tree.]{\includegraphics[width=0.32\linewidth]{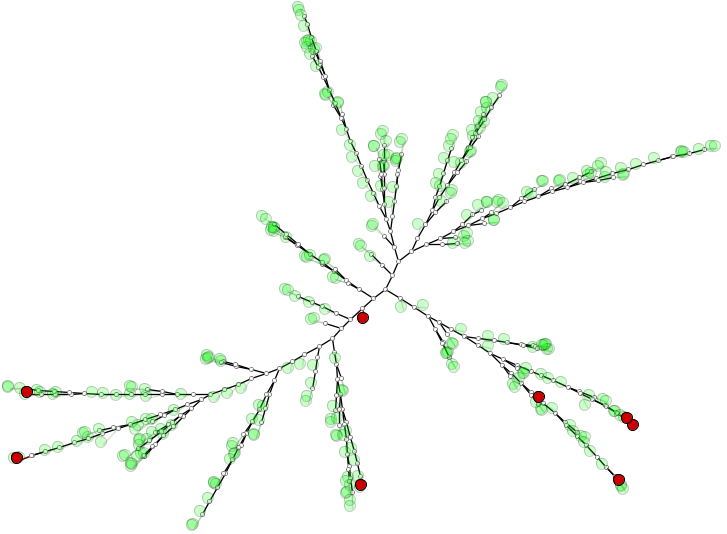}%
\label{fig_corel300_instProtoClassMatch}}
\hfil

\subfloat[Classification result.]{\includegraphics[width=0.32\linewidth]{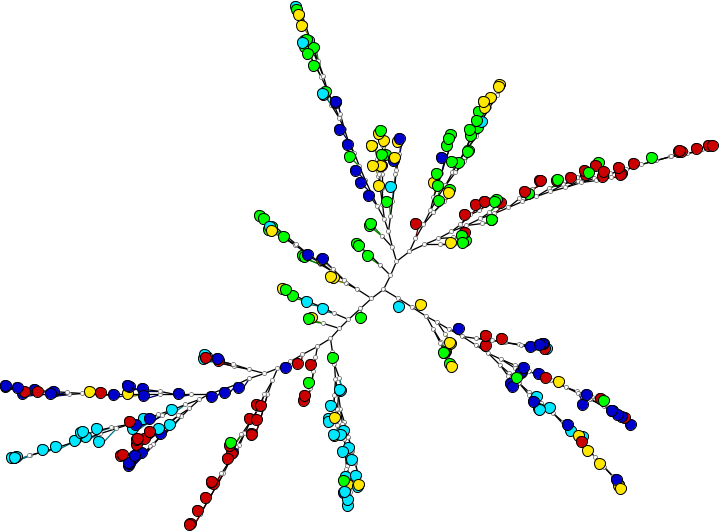}%
\label{fig_corel300_classResult}}
\hspace{0.1mm}
\subfloat[\textit{ClassMatch} tree.]{\includegraphics[width=0.32\linewidth]{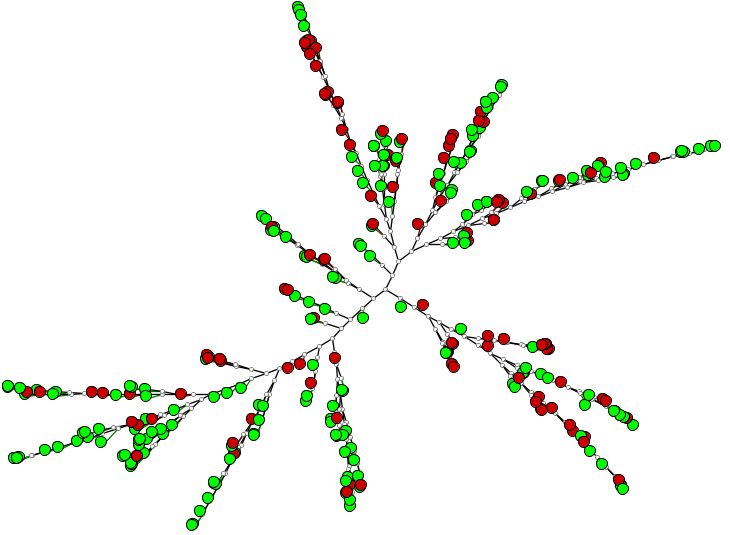}%
\label{fig_corel300_classMatch}}
\caption{MILTree's Bag Space Layout for the Corel-300 dataset. Visualization of the classification process from training set selection to classification result inspection.  Visualization of ground truth of dataset (a),  selected training (b),  \textit{InstancePrototypes ClassMatch} tree where bags with unsuitable instance prototypes are identified (c), visualization of classification result (d) and its correspondent \textit{ClassMatch} tree (e). Note that the \textit{InstancePrototypes ClassMatch} tree only shows the bags that were misclassified in the training set, whereas the \textit{ClassMatch} tree shows the bags that were misclassified in the test data and training set. Hence, for evaluating the classification results, users should only inspect the \textit{ClassMatch} tree (e). }
\label{corel300_MILTree}
\end{figure}

Here, we show the impact of adding new instances from bags that already exist in the training set to update and improve the classification model. Thus, in the updated model, some bags can be represented by more than one instance prototype. To update the model, the MILTree layout was used only in the bag space, and the MILTree-SI automatic selection method was used to detect new prototypes. We tested our approach with the multiclass Corel-300 dataset, containing 5 classes, 300 bags and 1293 instances. Figure \ref{fig_corel300_ground} displays the Corel-300 dataset in the bag space projection, where bags are points and the different colors represent different classes. 

As in the previous case study, an initial training set was selected using MILTree (see Figure~\ref{corel300_MILTree}b) to build an initial classifier for the Corel-300 dataset. We then show the \textit{InstancePrototypes ClassMatch} tree to identify bags that have unsuitable instance prototypes, as shown in Figure~\ref{corel300_MILTree}c. Finally, we add all the alternative prototypes automatically detected by the MILTree-SI selection method to the current model. Thus, the instances $B_{iy}$ for all red bags (see Figure~\ref{corel300_MILTree}c) are added to the classification model. 

Note again that in this case study, the new instance prototypes are added to the previously developed model. This strategy can be useful in multiclass problems because with more classes it can be difficult to select a single instance to represent each bag. The MILTree-SI method ranks the instances, and therefore, $B_{ix}$ and $B_{iy}$ are considered, respectively, the best and second-best prototypes so that it is intuitive to add the second prototype to the model. In contrast, the MILTree-Med method uses clustering in an attempt to obtain a pair of positive and negative prototypes, so it is highly recommended to choose just one of them because adding both instances to the model could increase the class overlap.

After updating the classifier, Figure~\ref{corel300_MILTree}d shows the classified MILTree, and Figure~\ref{corel300_MILTree}e shows its correspondent \textit{ClassMatch} tree, where green and red points represent bags correctly classified and misclassified, respectively. The accuracy before the update was 82.6\%, and after, including only eight new instances, it increased to 83.8\%. It is also worth mentioning that the achieved accuracy using a standard classification method, without our visual mining tool, was 78\%. This demonstrates that the selection of a proper training set and the process of updating the model using the proposed methods can help to create a classifier with improved performance, even in multiclass scenarios.

\subsection{Case 3: Adding New Bags Using the MILTree Visualization}

\begin{figure}[H]
\centering
\subfloat[Ground truth.]{\includegraphics[width=0.32\linewidth]{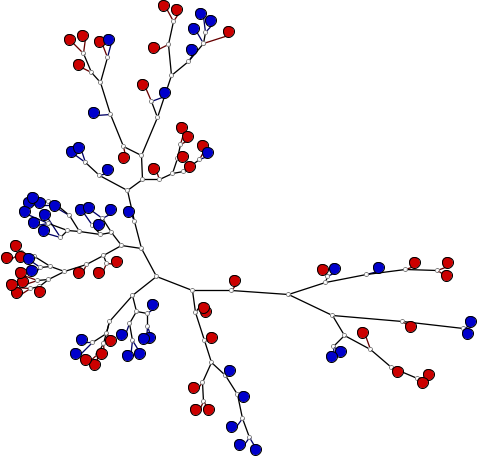}%
\label{fig_musk1_ground}}
\hfil
\hspace{0.2mm}
\subfloat[Selected training set.]{\includegraphics[width=0.32\linewidth]{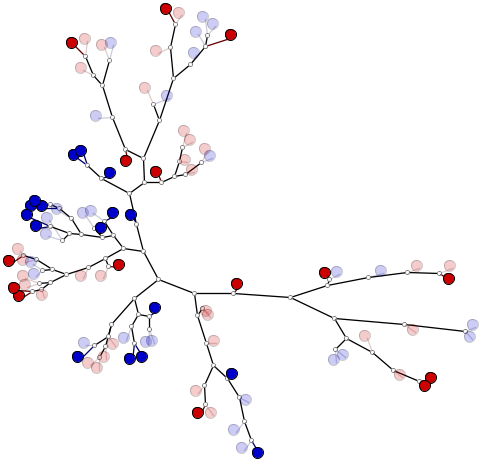}%
\label{fig_musk1_training}}
\hspace{0.1mm}
\subfloat[\textit{InstancePrototypes ClassMatch} tree.]{\includegraphics[width=0.32\linewidth]{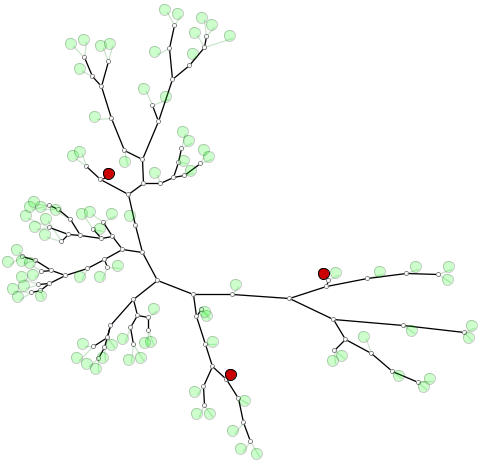}%
\label{fig_musk1_instprotoclassmatch}}
\hfil

\subfloat[\textit{ClassMatch} tree of initial classification result.]{\includegraphics[width=0.32\linewidth]{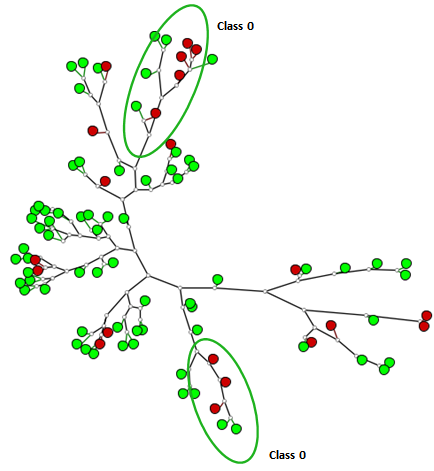}%
\label{fig_musk1_classMatchInitialClassification}}
\hspace{0.2mm}
\subfloat[Misclassified bags selected in the Ground truth.]{\includegraphics[width=0.32\linewidth]{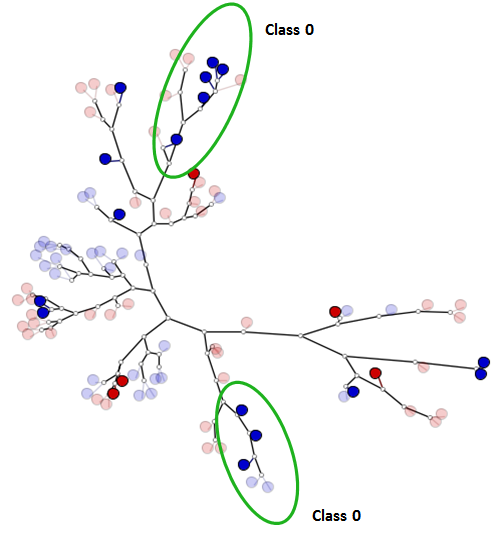}%
\label{fig_musk1_selectMisclassifiedbags}}
\hspace{0.4mm}
\subfloat[Selection of new bags located in the branches where bags were misclassified.]{\includegraphics[width=0.32\linewidth]{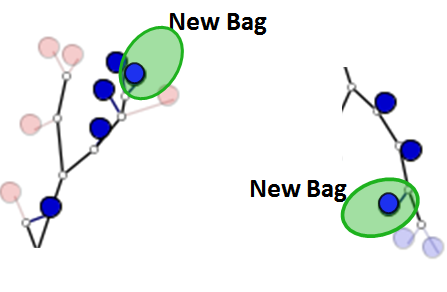}%
\label{fig_musk1_selectMisclassifiedbagsGround}}

\hfil
\subfloat[Classification result using updated model with new instance prototypes and new bags.]{\includegraphics[width=0.32\linewidth]{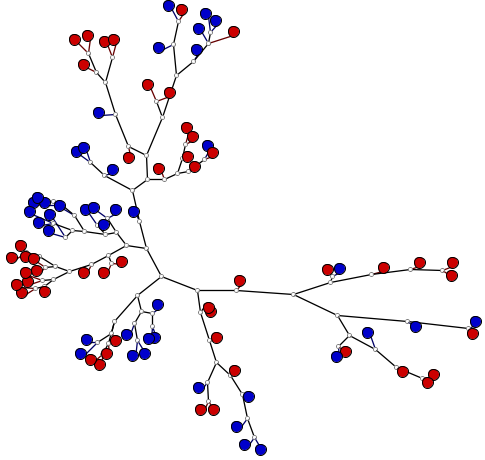}%
\label{fig_musk1_finalClassification}}
\hspace{0.5mm}
\subfloat[\textit{ClassMatch} tree of final classification result.]{\includegraphics[width=0.32\linewidth]{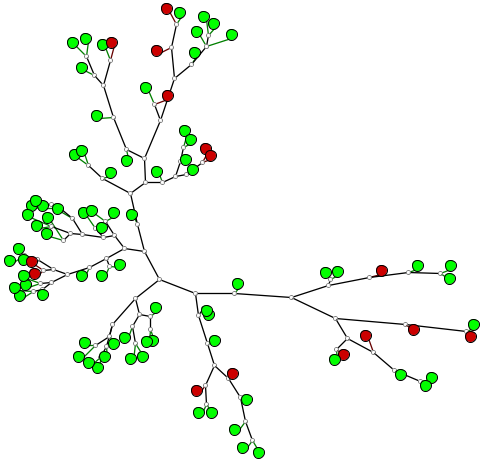}%
\label{fig_musk1_classMatchFinalClassification}}
\caption{Visualization of the classification process for the Musk1 dataset.}
\label{musk1_classification}
\end{figure}

In this case study, we illustrate how to use MILtree to identify and select both new prototypes and new bags to update a model and improve its performance, assuming that the initial training set is not representative enough to build a classifier with good class discrimination capabilities. The \textit{ClassMatch} tree is generated based on the initial classification results to identify misclassified bags, 
%MILTree for the projection 
and MILTree-SI is used as the instance prototype selection method. Similar to the second case study, we will only make use of the bag space layout of MILTree. We tested our system using the Musk1 benchmark dataset consisting of 92 bags and 476 instances. Figure~\ref{musk1_classification}a shows the projection of the Musk1 dataset in the bag space layout of MILTree, in which blue bags represent negative bags and red bags represent positive bags.

First, an initial training set was selected using MILTree (see Figure \ref{musk1_classification}b) to build the first classifier, and then the \textit{InstancePrototypes ClassMatch} tree is generated to highlight the misclassified bags on the training set, as shown in Figure~\ref{musk1_classification}c. Just like the second case study, the alternative instance prototypes $B_{iy}$ of all red bags (misclassified bags) are added to the initial model. In addition, we use the \textit{ClassMatch} tree to show the classifier results in the validation set to identify possible new bags that can be used to update the model as well. This is an interesting strategy because the validation set contains bags that were unseen in the training stage. Figure~\ref{musk1_classification}d presents the \textit{ClassMatch} tree obtained with the first classification result for the Musk1 dataset. 

Users can employ different strategies to keep updating the classification model using our layout by searching for representative bags in branches with high error rates and including those new bags in the training set. Instead of randomly selecting those bags, the visualization helps the user to identify subspaces, or regions of the feature space, that are poorly represented in the training set by looking at the tree branches with a high error rate, such as those annotated with ellipses in Figure~\ref{musk1_classification}d. 

\begin{figure}[H]
\centering
\includegraphics[width=0.35\textwidth]{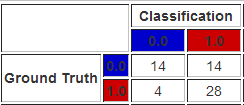}
\caption{Confusion matrix for the Musk1 dataset for classification results after using the initial classification model. Blue color represents the negative class, and the red color represents the positive~class.}
\label{fig_matrizConfusaoMusk1}
\end{figure}

The user can then assess results with the confusion matrix of the validation set. In our example, around half the negative bags (blue) were confused as positive (red), as shown in Figure \ref{fig_matrizConfusaoMusk1}. The layout shows a concentration of those errors in branches belonging to the same class and located in different regions of the tree; in other words, they are not neighbors in the tree projection, as shown in Figure \ref{musk1_classification}e. This may indicate that this class covers a wide range of features and may contain subclasses.

After inspecting and analyzing the \textit{ClassMatch} tree layout, the user selects new bags and updates the initial model. Figure \ref{musk1_classification}f highlights the two bags that were included in the model building process. Figure~\ref{musk1_classification}g shows the classified MILTree of the Musk1 test dataset, and Figure \ref{musk1_classification}h presents its correspondent \textit{ClassMatch} tree after the last model update. The accuracy on the Musk1 test dataset using the initial model was 73.9\%; after updating it using new instance prototypes, we achieved 75.2\%; and finally, after a second update using new bags, we obtained a 83.2\% accuracy. This demonstrates once again that visual mining selection strategies are helpful in the MIL scenario. 

\textls[-15]{To evaluate the performance of the proposed methods when compared to state-of-the-art MIL methods, Section \ref{sec:experiments} presents quantitative experiments performed over different~datasets.}

\vspace{-6pt}

\section{Experiments and Results}
\label{sec:experiments}
This section presents the experiments carried out to compare the proposed methods with MIL methods available for each dataset. We evaluate the average precision, average recall and average accuracy of both MILTree-Med and MILTree-SI using the MILTree layout for five  MIL benchmark datasets, Corel-1000 and Corel-2000 image classification datasets, as well as the Biocreative text classification dataset. Furthermore, we perform experiments on a large-scale dataset and in a multiclass problem, both not addressed in previous works. The proposed methodology called MILSIPTree is used to carry out the complete multiple-instance classification process. Note that MILSIPTree is supported by both MILTree-Med and MILTree-SI instance prototype selection methods and the MILTree layout as well.

The source code of the proposed methods and the multi-instance datasets are publicly available to allow reproducibility. The LIBSVM package was applied to train all SVMs using settings that are comparable with the results obtained by the competing methods. For the Musk1 and Musk2 datasets, we have employed the classifier \textit{nu-SVC (Nu-Support Vector Classification)}, with a \textit{Nu} value equal to 0.6. For the images datasets (Elephant, Fox, Tiger, Corel-1000 and Corel-2000) and text dataset (Biocreative), we used the classifier \textit{C-SVC (C-Support Vector Classification)}, with a \textit{Cost} value equal to 1. No kernel was used because we intended to show how the proposed methods help obtaining a feature space with linearly separable classes, which require less effort on designing the classifier.

\subsection{Benchmark Datasets}
Five standard MIL benchmarks were used: The Musk1 and Musk2 datasets~\cite{Dietterich1997}, as well as Elephant, Fox and Tiger image datasets~\cite{Andrews2003supportvector}. Those have been widely used in multiple-instance learning studies.

Musk1 and Musk2 are real-world benchmark datasets available at the UCI machine learning repository~\cite{Lichman2013}. The Musk data were generated in the research of drug activity prediction, in which a drug molecule is represented by a bag, and the different structural conformations of this molecule are considered as instances. Musk1 contains 47 positive bags and 45 negative bags, and the number of instances contained in each bag ranges from 2 to 40. Musk2 contains 39 positive bags and 63 negative bags, and the number of instances contained in each bag ranges from 1 to 1044. Each instance is represented by 166 continuous attributes. Table \ref{table:benchmarkMusk} shows the detailed information on the Musk datasets.

The image datasets named Elephant, Fox and Tiger were built with the goal of discriminating images containing elephants, foxes and tigers from those that do not, respectively. In this case, bags are considered images, and instances are considered regions of interest within the images. More details about these datasets are given in Table \ref{table:benchmarkImage}. 

We split each dataset into 30\% training and 70\% testing data. Our methods allow the training set to be small because they provide smart ways to select it so that it is representative enough. The model update was similar to the case studies 1 and 2: the initial model is updated by either changing the prototypes using MILTree-Med or including new prototypes using MILTree-SI. 

Table \ref{table:benchmarkMILTree} compares the results achieved by MILTree-SI and MILTree-Med in detail. Additionally, in Table \ref{table:benchmarkCompare}, we compare MILTree-SI and MILTree-Med with the accuracy reported by nine MIL algorithms from the literature: Four baseline methods such as EM-DD \cite{Zhang2001EMDD}, DD-SVM \cite{Chen2004DDSVM}, mi-SVM\cite{Andrews2003supportvector} and MI-SVM\cite{Andrews2003supportvector}, and five methods that use instance selection such as MILES \cite{Yixin2006MILES}, MILIS \cite{Fu2011MILIS}, MILD-B \cite{Li2010MILD} as well as the most recent state-of-the-art results from MILSIS \cite{Huang2012SalientInst} and MILDE \cite{Amores2015MILDE}. The best accuracies are shown in~bold.

The actual number of bags used to update the initial model was between three and eight. These bags were selected through the visual analysis of the data and by choosing alternative prototypes that were automatically detected by using either SI or Med approaches. The results show that the proposed MILTree-SI and MILTree-Med methods are very competitive, especially MILTree-Med, achieving an overall average performance of 82.8\%.

\begin{specialtable}[H] 
\caption{Musk datasets and the average number of instances per bag(Inst/Bag) for each dataset.\label{table:benchmarkMusk}}
\begin{tabularx}{\linewidth}{>{\centering\arraybackslash}X>{\centering\arraybackslash}X>{\centering\arraybackslash}X>{\centering\arraybackslash}X>{\centering\arraybackslash}X>{\centering\arraybackslash}X}
\toprule
&\multicolumn{2}{c}{\textbf{Bags}}&\multicolumn{2}{c}{\textbf{Instances}}  &\\
\cmidrule{2-5}
\textbf{Dataset}&\textbf{Total}&\textbf{Pos./Neg.}& \textbf{Total}& \textbf{Min/Max}& \textbf{Dim} \\
\midrule
Musk1&92&47/45& 476& 2/40& 166 \\
Musk2&102&39/63& 6598& 1/1044& 166 \\
\bottomrule
\end{tabularx}
\end{specialtable}\vspace{-6pt}

\begin{specialtable}[H] 
\caption{Image datasets and the average number of instances per bag(Inst/Bag) for each dataset.\label{table:benchmarkImage}}
\begin{tabularx}{\linewidth}{>{\centering\arraybackslash}X>{\centering\arraybackslash}X>{\centering\arraybackslash}X>{\centering\arraybackslash}Xc>{\centering\arraybackslash}X}
\toprule
&\multicolumn{2}{c}{\textbf{Bags}}&\multicolumn{2}{c}{\textbf{Instances}}  &\\
\cmidrule{2-5}
\textbf{Dataset}&\textbf{Total}&\textbf{Pos./Neg.}& \textbf{Total}& \textbf{Avg. Inst./Bag}& \textbf{Dim} \\
\midrule
Elephant&200&100/100& 1391& 6.96& 230 \\
Fox&200&100/100&1220& 6.10& 230 \\
Tiger&200&100/100& 1320& 6.60& 230 \\
\bottomrule
\end{tabularx}
\end{specialtable}\vspace{-6pt}

\begin{specialtable}[H]
\caption{Results of classification using MILTree-Med and MILTree-SI on the benchmark datasets\label{table:benchmarkMILTree}}
\begin{tabularx}{\linewidth}{c>{\centering\arraybackslash}X>{\centering\arraybackslash}X>{\centering\arraybackslash}X>{\centering\arraybackslash}X>{\centering\arraybackslash}X>{\centering\arraybackslash}X>{\centering\arraybackslash}X>{\centering\arraybackslash}X}
\toprule
&\multicolumn{4}{c}{\textbf{MILTree-Med}} & \multicolumn{4}{c}{\textbf{MILTree-SI}}  \\
\cmidrule{2-9}
\textbf{Dataset} &\textbf{Accur} &\textbf{Prec}& \textbf{Recall}& \textbf{F1} &\textbf{Accur} &\textbf{Prec}& \textbf{Recall}& \textbf{F1}\\
\midrule
Musk1 &\textbf{83.2} &83.23 &81.7 &0.82 &\textbf{83.2} &82.4 &81.7 &0.82 \\
Musk2 &\textbf{91.8} &91.4 &91.4 &0.91 &85.4 &84.4 &84.3 &0.84 \\
Elephant &\textbf{83.1} &81.7 &81.6 &0.82 &81.4 &79.4 &79.4 &0.79 \\
Fox   &\textbf{72.7} &68.3 &68.3 &0.68 &\textbf{72.7} &68.3 &68.3 &0.68 \\
Tiger &\textbf{83.0} &82.0 &81.4 &0.82 &82.9 &83.4 &81.4 &0.82 \\
\bottomrule
\end{tabularx}
\end{specialtable}
\vspace{-6pt}
\begin{specialtable}[H]
\caption{Comparison between MILTree-SI / MILTree-Med and related methods from the literature on the benchmark datasets.\label{table:benchmarkCompare}}
\begin{tabularx}{\linewidth}{c>{\centering\arraybackslash}X>{\centering\arraybackslash}X>{\centering\arraybackslash}X>{\centering\arraybackslash}X>{\centering\arraybackslash}X>{\centering\arraybackslash}X}
\toprule
\textbf{Method} &\textbf{Musk1} &\textbf{Musk2} &\textbf{Elephant} &\textbf{Fox}  &\textbf{Tiger} &\textbf{Avg.}\\
\midrule
MILTree-Med &83.2 &\textbf{91.8} &83.1 &\textbf{72.7} &\textbf{83.0} &\textbf{82.8}\\
MILTree-SI &82.3 &85.4 &81.4 &72.7 &82.9 &81.1\\
EM-DD &84.8 &84.9 &78.3 &56.1 &72.1 &75.2 \\
MI-SVM &77.9 &84.3 &73.1 &58.8 &66.6 &72.1 \\
mi-SVM &87.4 &83.6 &80 &57.9 &78.9 &77.6\\
DD-SVM &85.8 &91.3 &83.5 &56.6 &77.2 &79.0\\
MILD-B &88.3 &86.8 &82.9 &55.0 &75.8 & 77.8\\
MILIS &88.6 &91.1 &- &- &- &-\\
MILES &86.3 &87.7 &84.1 &63.0 &80.7 &80.4\\
MILSIS &\textbf{90.1} &85.6 &81.8 &66.4 &80.0 &80.9 \\
MILDE &87.1 &91.0 &\textbf{85} &66.5 &83.0 &82.5 \\
\bottomrule
\end{tabularx}
{\footnotesize Bold values indicate the method that obtains the best performance for each dataset.}
\end{specialtable}

\subsection{Image Classification}
\label{subsec:imageclass}
We used the Corel-1000 image dataset to evaluate the performance of MILTree-Med and MILTree-SI using the MILTree layout for analyzing multi-instance data. It contains 10 subcategories representing distinct topics of interest. Each subcategory contains 100~images. 

Since the original Corel dataset was not designed for multi-instance learning, we adopt the same approach of \cite{Chen2004DDSVM} and~\cite{Yixin2006MILES} to segment each image into several regions. The second column of Tables~\ref{table:imageClassMED} and~\ref{table:imageClassSI} presents the average number of instances per bag for each category. We then choose one category as the positive class and select 100~images uniformly from the remaining categories to create the negative class as is performed in~\cite{Andrews2003supportvector}. The same process is followed for each category, totaling 10 subsets of categories. 
In this section, each subdataset is split into 20\% training and 80\% testing data. 

The classification accuracy, precision and recall rates for both experiments are reported in Tables \ref{table:imageClassMED} and \ref{table:imageClassSI}. In both tables, \textit{Proto} represents the number of instance prototypes that were updated in the bag or instance space layouts. \textit{AddBags} represents the number of bags that were included in the training data from the bag space layout. \textit{AddProto} represents the number of new instances $B_{iy}$ included from the bags belonging to training data, which represents the MILTree-SI approach. Recall this was designed to identify additional prototypes to be included in the current model to reinforce the representation of misclassified bags.

\begin{specialtable}[H]
\caption{ Classification results using MILTree-Med on the Corel Dataset.\label{table:imageClassMED}}
\begin{tabularx}{\linewidth}{c>{\centering\arraybackslash}X>{\centering\arraybackslash}X>{\centering\arraybackslash}X>{\centering\arraybackslash}X>{\centering\arraybackslash}Xc>{\centering\arraybackslash}X}
\toprule
&&\multicolumn{4}{c}{\textbf{Measures}}  & &\\
\cmidrule{3-6}
\textbf{Category ID} &\textbf{Inst/Bag}&\textbf{Accur} &\textbf{Prec}& \textbf{Recall}& \textbf{F1}& \textbf{Proto}& \textbf{AddBags}\\
\midrule
Category0 &4.84 & 75.96& 72.68 &72.67  &0.73  & 7& 0\\
Category1 &3.54 & 79.07& 78.74 & 76.74 & 0.78 & 3&2\\
Category2 &3.1 &78.97 &79.16  & 76.67 & 0.78 & 6&3\\
Category3 &7.59 & 91.25&90.89 & 90.85 & 0.91 & 4&0\\
Category4 &2.00 & 78.4& 79.68 & 75.95 & 0.78 &4 &0\\
Category5 &3.02 &81.91 & 83.21 & 80.26 & 0.82& 4 &1\\
Category6 &4.46 &89.09 &  88.68& 88.49 & 0.88 & 1 &0\\
Category7 &3.89 & 84.23& 83.62 & 82.91 & 0.83 & 6 &2\\
Category8 &3.38 & 81.19& 79.46 & 79.25 &0.79  & 1 &1\\
Category9 &7.24 & 81.26& 81.26 &79.39  & 0.8 & 1 &0\\
\bottomrule
\end{tabularx}
\end{specialtable}\vspace{-6pt}
\begin{specialtable}[H]
\caption{Classification results using MILTree-SI on the Corel Dataset.\label{table:imageClassSI}}
\begin{tabularx}{\linewidth}{c>{\centering\arraybackslash}X>{\centering\arraybackslash}X>{\centering\arraybackslash}X>{\centering\arraybackslash}X>{\centering\arraybackslash}Xc>{\centering\arraybackslash}X>{\centering\arraybackslash}X}
\toprule
&&\multicolumn{4}{c}{\textbf{Measures}}  & &&\\
\cmidrule{3-6}
\textbf{Category ID} &\textbf{Inst/Bag}&\textbf{Accur} &\textbf{Prec}& \textbf{Recall}& \textbf{F1}& \textbf{Proto}& \textbf{AddProto}& \textbf{AddBags}\\
\midrule
Category0 &4.84 &68.14 & 62.24 & 62.11& 0.62 & 0& 11& 2\\
Category1 &3.54 &75.82& 74.94 & 72.67 & 0.74 &3 & 0 &0 \\
Category2 &3.1 & 76.47& 73.33 &73.33 &0.73  &3 & 0 &2 \\
Category3 &7.59  &72.88 & 68.94 &68.63 & 0.69 & 0& 12&0\\
Category4 &2.00  & 83.65& 84.99 & 82.28& 0.84 &4& 0& 1\\
Category5 &3.02  & 80.86&  80.84& 78.95& 0.8 & 0 &14& 0\\
Category6 &4.46  & 89.09& 88.58 & 88.46& 0.89 & 1& 0& 1\\
Category7 &3.89  & 80.03& 78.28 &77.85 & 0.78 & 0& 6& 0\\
Category8 &3.38  &75.21 & 71.7 & 71.7& 0.72 & 0& 0 &2 \\
Category9 &7.24  &75.74 &  72.39&72.39 & 0.72 & 0&6 & 0\\
\bottomrule
\end{tabularx}
\end{specialtable}

Table \ref{table:imageClassComparer} presents the accuracy of different methods from the literature, including EM-DD~\cite{Zhang2001EMDD}, mi-SVM~\cite{Andrews2003supportvector}, MI-SVM~\cite{Andrews2003supportvector}, DD-SVM~\cite{Chen2004DDSVM} and SMILES~\cite{Xiao2014SmileFramework}. We can see that our MILTree-SI and MILTree-Med methods achieve high classification accuracy on the sub-datasets. In particular, MILTree-Med outperforms all the others in all but three datasets. The competing methods EMDD, MI-SVM and DD-SVM selects only one instance as a prototype, which is often not sufficient. Moreover, our method was able to be competitive when compared with mi-SVM and SMILES, even though both use all instances from each bag to build the classifier.

\subsection{Multiple-instance Multiclass Datasets}
\label{sec:multiclass}

In this section, we turn our attention to the performance of MILTree-Med and MILTree-SI using the MILTree layout for solving multiclass classification problems. The baseline methods, such as EM-DD, mi-SVM, MI-SVM and DD-SVM, were originally proposed for binary class classification. Our MILTree layout and MILTree-Med and MILTree-SI methods also support multiclass datasets. We extend MILTree-Med and MILTree-SI for multiclass by performing one-against-all by decomposing the problem into a number of binary classifiers that are created to separate each class from the remaining ones.

\begin{specialtable}[H] 
\caption{{Comparison between} MILTree-SI/MILTree-Med and related methods from the literature on the Corel Dataset.\label{table:imageClassComparer}}%MDPI: Please confirm if the bold on those figures are necessary. if it does, please define.
\begin{tabularx}{\linewidth}{>{\centering\arraybackslash}X>{\centering\arraybackslash}Xccc>{\centering\arraybackslash}Xcc}
\toprule
\textbf{Datasets} &\textbf{EMDD} &\textbf{mi-SVM} &\textbf{MI-SVM} &\textbf{DD-SVM}  &\textbf{SMILES} &\textbf{MILTree-SI} &\textbf{MILTree-Med}\\
\midrule
Cat0 &68.7 &71.1 &69.6 &70.9	 &72.4 &68.1 &\textbf{76.0} \\
Cat1 &56.7 &58.7 &56.4 &58.5	 &62.7 &75.8 &\textbf{79.1} \\
Cat2 &65.1 &67.9 &66.9 &68.6	 &69.6 &76.5 &\textbf{79.0} \\
Cat3 &85.1 &88.6 &84.9 &85.2	 &90.1 &72.9 &\textbf{91.3} \\
Cat4 &96.2 &94.8 &95.3 &\textbf{96.9}	 &96.6 &83.7 &78.4 \\
Cat5 &74.2 &80.4 &74.4 &78.2	 &80.5 &80.9 &\textbf{81.9} \\
Cat6 &77.9 &82.5 &82.7 &77.9	 &83.3 &89.1 &\textbf{89.1} \\
Cat7 &91.4 &93.4 &92.1 &94.4	 &\textbf{94.7} &80.3 &84.2 \\
Cat8 &70.9 &72.5 &67.2 &71.8	 &73.8 &75.2 &\textbf{81.2} \\
Cat9 &80.2 &84.6 &83.4 &84.7	 &\textbf{84.9} &75.8 &81.3 \\
\bottomrule
\end{tabularx}
{\footnotesize Bold values indicate the method that obtains the best performance in each dataset.}
\end{specialtable}

\begin{figure}[H]
\centering
% width=0.9
\includegraphics[width=0.7\textwidth]{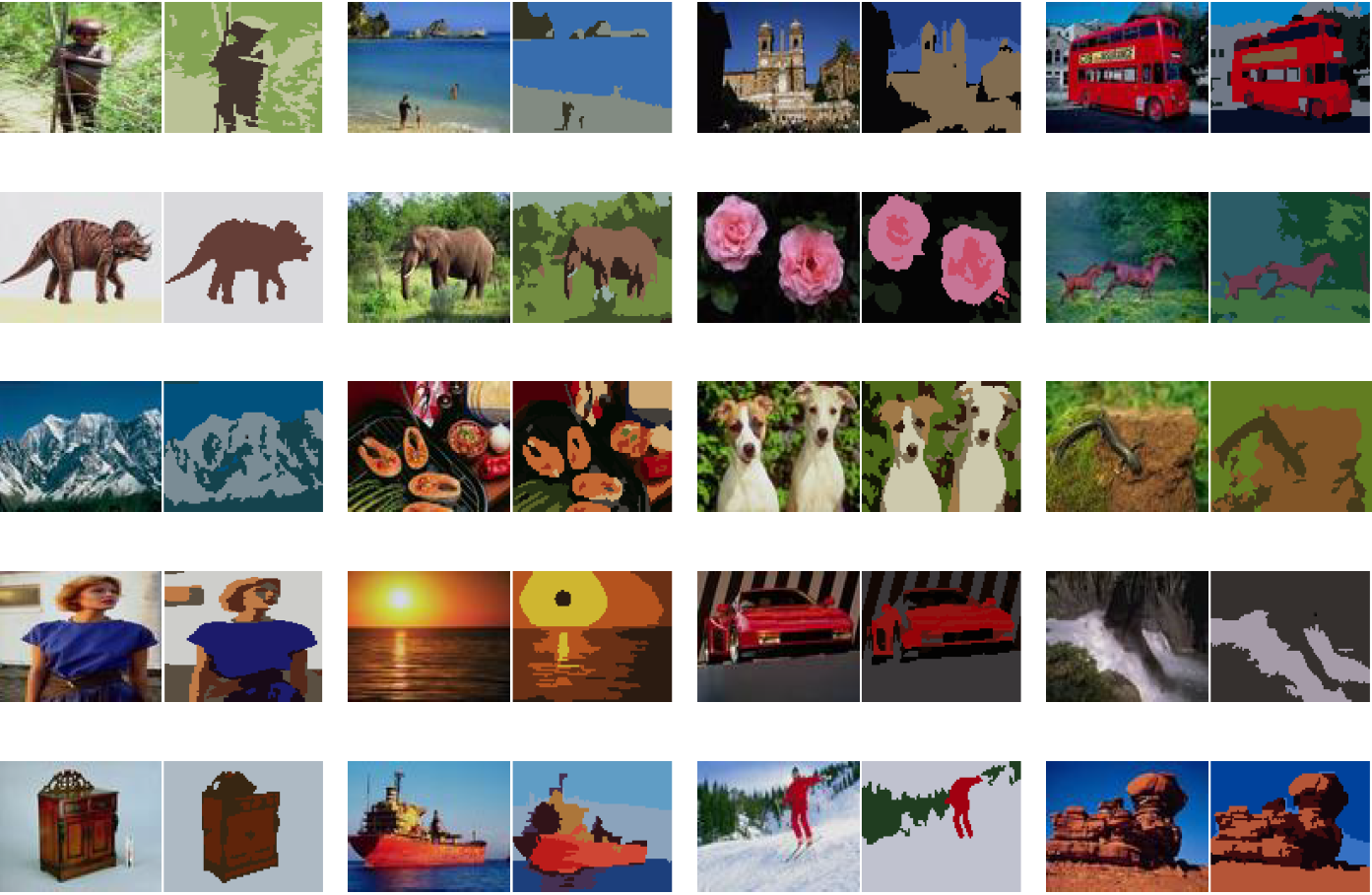}
\caption{Images randomly sampled from 20 categories of the COREL dataset and the corresponding segmentation results. Segmented regions are shown in their representative colors.}
\label{fig_corel2000}
\end{figure}

We used the Corel-2000 dataset with 2000 images, 20 classes and 100 images per class. Details about segmentation and feature extraction were mentioned in Section~\ref{subsec:imageclass}. Two experiments were carried out: one using the first 10 categories in the dataset (Corel-1000), and a second one using the complete dataset with all 20 categories (Corel-2000). Figure \ref{fig_corel2000} shows images randomly sampled from the 20 categories.%Please check and revise. % authors:  checked 
%Authors: The missing figure was added.

\begin{specialtable}[H]
\caption{Comparison between MILTree-SI / MILTree-Med and related methods from the literature on the 1000-Corel and 2000-Corel Datasets.\label{table:CorelCompare}}
\begin{tabularx}{\linewidth}{>{\centering\arraybackslash}X>{\centering\arraybackslash}X>{\centering\arraybackslash}X}
\toprule
\textbf{Method} &\textbf{1000-Corel} &\textbf{2000-Corel }\\
\midrule
MILTree-Med &\textbf{93.1} &\textbf{93.9} \\
MILTree-SI &90.3 &93.9\\
MI-SVM &75.1 &54.6  \\
mi-SVM &76.4 &53.7 \\
DD-SVM &81.5 &67.5\\
MILIS &83.8 &70.1 \\
MILES &82.3 &68.7 \\
MILDE &- &74.8 \\
\bottomrule
\end{tabularx}
{\footnotesize Bold values indicate the method that obtains the best performance in each dataset.}
\end{specialtable}

Table \ref{table:CorelCompare} presents the classification accuracy rates, including the results of DD-SVM, MILES and MILIS, as reported by the original papers, and the results of MI-SVM and mi-SVM, as reported in~\cite{Fu2011MILIS}. From Table \ref{table:CorelCompare}, we can see that MILTree-Med and MILTree-SI outperform competing methods due to the efficient bag selection strategy used for training and the efficient instance prototype selection performed inside each bag. 

\subsection{Scalability Analysis on a MIL Text Classification}
\label{subsec:corel}
In MIL text classification, each document is represented as a bag and the document paragraphs as instances. The Biocreative dataset used in this experiment has 1623 documents (papers) extracted from biomedical journals, belonging to three text categories: Components, Processes and Functions, all referring to Gene Ontologies (GOs)~\cite{Ray2005supervisedversus}. It contains 34,569 instances, posing a challenge to conventional visualizations. Table~\ref{tab:textdataset} details this dataset. Each text document in the collection has a Protein identification, an associated article ID in PUBMED and a description text.

\begin{specialtable}[H]
\caption{Biocreative dataset. Total number of bags and instances for each category.\label{tab:textdataset}}
\begin{tabularx}{\linewidth}{>{\centering\arraybackslash}X>{\centering\arraybackslash}X>{\centering\arraybackslash}X>{\centering\arraybackslash}X}
\toprule
\textbf{Dataset} &\textbf{Bags} &\textbf{Instances} &\textbf{Dimensions} \\
\midrule
Components &423 &9104 &200 \\
Functions & 443 & 9387 &200 \\
Processes & 757 & 25181 &200 \\
\midrule
Total & 1623 & 34569 &600 \\
\bottomrule
\end{tabularx}
\end{specialtable}

We split the dataset in about 10\% training and 90\% testing data. After selecting the training set with MILTree and using the proposed prototype selection methods, all training bags were correctly classified, which means that all instances chosen as instance prototypes by MILTree-SI and MILTree-Med selection methods were representative.

To compare our methods with other state-of-the-art methods, we employed the Weka machine learning package ({\url{http://www.cs.waikato.ac.nz/ml/weka}}) (accessed on 20 August 2021). %MDPI: footnote is not allowed in our journal. Please include this paragraph to the maintext. Please add the access date as Day Month Year.
%Authors: Access date was added.
There are no previous results for this dataset employing the aforementioned methods, such as MILES, SMILES and others, for the three categories of Biocreative dataset. Previous work only shows results for one category, such as~\cite{Wei2014ScalableMIL}, that presents result only for the ``Process'' category.  For this reason, we compare our methods with multi-instance methods available in Weka, such as DD, EM-DD, MI-SVM, MIWrapper~\cite{Frank2003MIWrapper}, TLDSimple~\cite{Xu2003statisticallearning} and MIBoost~\cite{Freund1996MIBoost}. Table~\ref{tab:BiocreativeCompare} shows the results, where both MILtree-Med and MILTree-SI methods supported by the MILTree layout obtained higher accuracy.

\begin{specialtable}[H] 
\caption{Comparison of classification accuracy between MILTree-SI/MILTree-Med and baseline methods on the Biocreative dataset.\label{tab:BiocreativeCompare}}
\begin{tabularx}{\linewidth}{>{\centering\arraybackslash}X>{\centering\arraybackslash}X}
\toprule
\textbf{Method} &\textbf{Biocreative} \\
\midrule
MILTree-Med &\textbf{99.1}\\
MILTree-SI &96.3\\
\midrule
MI-SVM &90.9  \\
EM-DD &91.0 \\
DD & 90.9\\
MIWrapper & 90.5\\
TLDSimple & 85.0\\
MIBoost & 90.5\\
\bottomrule
\end{tabularx}
{\footnotesize Bold value indicates the method that obtains the best performance.}
\end{specialtable}

\begin{figure}[H]
\centering
\subfloat[]{\includegraphics[width=1.8in,height=2.0in]{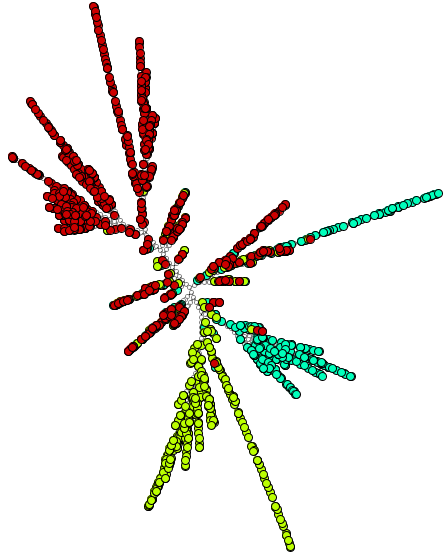}%
\label{biocreativegroundtruth}}
\hfil
\hspace{2.1mm}
\subfloat[]{\includegraphics[width=1.8in,height=2.0in]{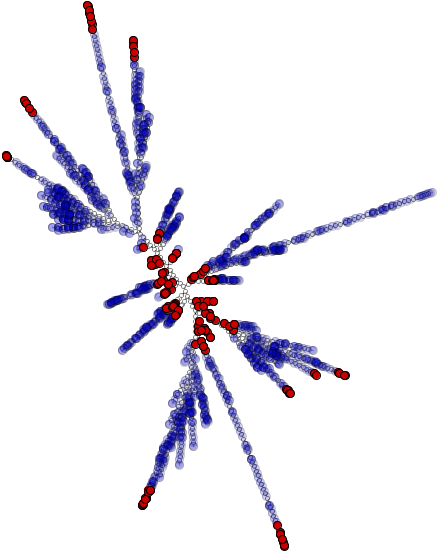}%
\label{biocreativetraining}}
\hspace{0.1mm}
\caption{Bag space projection of MILTree for the Biocreative dataset using MILTree-SI, with the projection of its \textit{ground truth} (\textbf{a}) and the selected training data (red bags) (\textbf{b}).}
\label{fig:biocreativedata}
\end{figure}

The results can be explained by the visual discrimination of the categories ``Components'', ``Processes'' and ``Functions'' in the bag space projection of MILTree (see \mbox{Figure \ref{fig:biocreativedata}a}), which provides a clear guideline for users when identifying representative bags for the training set. This corroborates previous results that favor the strategy of selecting samples from both the internal and external parts of the MILTree. In Figure~\ref{fig:biocreativedata}b, we show the training sample selected following the established guidelines.

\subsection{MILTree Layout Bag Positioning}
\label{sec:ImpactBagsPositioning}

This experiment aims at evaluating how the bag positions in the MILTree layout are related to good candidates for training set selection. As mentioned in Section~\ref{sec:MILSIPTreeInstSpaceProj}, the MILTree projects a bag belonging to a given class as an external point of the tree if it is furthest from the remaining classes. 
At the core of the MILTree (internal points) will be the bags that are closest to other classes, as well as the ones that overlap in feature space.

We followed a similar methodology to investigate the impact of bag positioning on the classification results. Three training sets are used in this analysis; the first training set is composed only of external instances, the second training set is composed only of internal instances and the third training set is composed of a combination of the first and second training sets.

The Corel's Cat3 and Cat6 subdatasets and MILTree-Med were used for this experiment. For Cat3, a total of 47 training bags are selected as training examples, while the remaining 153 bags are used as test set. For the Cat6, 44 training bags were selected for training, while the remaining 156 bags are used as the test set. Table \ref{tab:TypesTrainingSets} shows the results of each multi-instance classification for both collections. When using just external points, the model is often unable to represent boundary elements, resulting in a classifier that does not take into account the overlap degree between the classes. Using only internal bags, we add this information in the training set, but by combining external and internal bags, we have a sample containing both the more class-distinct elements and the ones belonging to the decision boundary region, resulting in a more accurate classifier. \\

\begin{specialtable}[H] 
\caption{Results of multi-instance classification using three types of training set.\label{tab:TypesTrainingSets}}
\begin{tabularx}{\linewidth}{>{\centering\arraybackslash}X>{\centering\arraybackslash}X>{\centering\arraybackslash}X>{\centering\arraybackslash}X}
\toprule
&&\textbf{Cat3}& \\
\midrule
&\textbf{External} &\textbf{Internal} &\textbf{Combined} \\
&\textbf{Bags} &\textbf{Bags} &\textbf{Bags} \\
\midrule
Matching Bags &104 (68\%) &134 (73.1\%) &137 (89.5\%) \\
Non-Matching Bags & 49(32\%) &19 (26.9\%) &16 (10.5\%) \\
Accuracy & 72.16\% & 88.31\% &90.06\% \\
Precision & 71.21\% & 87.74\% &89.58\% \\
Recall & 67.97\% & 87.58\% &89.54\% \\
\midrule
&&\textbf{Cat6}& \\
\midrule
&\textbf{External} &\textbf{Internal} &\textbf{Combined} \\
&\textbf{Bags} &\textbf{Bags} &\textbf{Bags} \\
\midrule
Matching Bags &134 (85.9\%) &114 (73.1\%) &139 (89.1\%) \\
Non-Matching Bags & 22(14.1\%) &42 (26.9\%) &17 (10.9\%) \\
Accuracy & 86.82\% & 76.28\% &89.66\% \\
Precision & 85.93\% & 73.07\% &89.28\% \\
Recall & 85.9\% & 73.08\% &89.1\% \\
\bottomrule
\end{tabularx}
\end{specialtable}

\section{User Study}
\label{sec:UsabilityTesting}

In this section, we present a user study to evaluate the usability of the MILTree layout for multiple-instance learning problems. We conducted the user study with five participants, three male and two female, who were all undergraduate or graduate students. The age of the participants ranged from 20 to 33. All participants, except one who received additional guidance, had previous knowledge about supervised learning and classification models. All users performed the same task, which was to build a multiple-instance model for the Corel-300 dataset (see Section~\ref{subsec:casetwo} for details).

\begin{specialtable}[H] 
\caption{The questionnaire used in the evaluation.\label{tab:questionnaire}}
\begin{tabularx}{\linewidth}{cX}
\toprule
\textbf{No.} &\textbf{Questions} \\
\midrule
1 &Is it possible to clearly identify the classes using MILTree?\\
2 &Does the browsing through MILTree (Bags and Instances projection  space) conduce the user to better understand the structure of multiple-instance data?\\
3 &Is the Prototypes ClassMatch tree useful for identifying misclassified bags? \\
4 &\textls[-10]{Is the ClassMatch tree useful for discovering new bags that help to improve or update the~model?} \\
5 & \textls[-20]{Do you feel that MILTree provides useful support in the multiple-instance classification~process?}\\
\bottomrule
\end{tabularx}
\end{specialtable}

% \begin{specialtable}[!htp] 
% \centering
% \caption{The questionnaire used in the evaluation.\label{tab:questionnaire}}
% \begin{tabular}{l l l }
% \toprule[0.1mm]
% \toprule[0.4mm]
% &&\\
% \textbf{No.} &\textbf{Questions} \\
% \midrule
% 1 &Is it possible to clearly identify the classes using MILTree?\\
% 2 &Does the browsing through of MILTree (Bags and Instances projection space) conduces the user to \\
% & better understand the structure of multiple-instance data?\\
% 3 &Is the Prototypes ClassMatch tree useful for identifying misclassified bags? \\
% 4 &Is the ClassMatch tree useful to discover new bags that help to improve or update the model ? \\
% 5 & Do you feel that MILTree provide useful support in the multiple-instance classification process?\\
% \bottomrule[0.4mm]
% \end{tabular}
% \end{specialtable}

We prepared two 10-minute long videos to instruct how to use MILTree for multiple-instance classification where we used datasets other than Corel-300 as examples. All participants watched the training videos prior to starting the study. We also introduced MILTree to all participants and showed how they could use it.
All participants used MILTree-SI as the instance prototype selection method and were instructed to use around 20\% of the data to train the model, while the remaining 80\% would be used for validation and testing.

After finishing the task, each participant was asked to answer a multiple choice questionnaire, with the questions shown in Table~\ref{tab:questionnaire}, and to justify the grade in a few sentences. 
For each question the answers 0, 1, 2, 3 and 4 were available, where 0 is the worst score and 4 is the best score. The grades had the following meanings: 0---No; 1---Little; 2---Fair; 3---Good; 4---Excellent. 
%Is the bold necessary? Please confirm.
%Authors: No, removed.

Table~\ref{tab:questionnaireResults} presents the means of the grades given to the questions. These results indicate that MILTree provides effective support to MIL by the subjects in the case study. 

\begin{specialtable}[H]
\caption{Means of the results obtained in the evaluation using the questionnaire for multi-instance classification of Corel-300 dataset.\label{tab:questionnaireResults}}
\begin{tabularx}{\linewidth}{>{\centering\arraybackslash}X>{\centering\arraybackslash}X}
\toprule
\textbf{Questions} &\textbf{Grade} \\
\midrule
Question 1 &3.2\\
Question 2 &2.8\\
Question 3 &3 \\
Question 4 &2.4 \\
Question 5 & 3\\
\bottomrule
\end{tabularx}
\end{specialtable}

All participants agreed that MILTree provided a good understanding of the multiple-instance data structure and supported them in the classification task. They had some comments as follows:
``\textit{Identifying the classes within dataset is very simple because, generally, instances of the same class are closer}''; ``\textit{In two simple steps it is possible to identify and update misclassified bags using Prototypes-ClassMatch tree}''; ``\textit{ClassMatch tree is useful because we can identify new bags that could be more representative for each class.}''

About MILTree as a tool for multi-instance classification, the majority of participants said that it is easy to use, leaving comments such as: 
``\textit{It is a very useful tool for some tasks with multiple-instance data, such as selection of training data, classification and updating of the training data. In these tasks, the tool is easy to use, differently of some other approaches that not use visualization of multi-instance data.}'' 

On the less positive side, two participants recommended that the browsing through of MILTree should be improved, leaving comments such as: 
``\textit{In some functionalities, you have to do many clicks to see the results. In some cases it is difficult to see what data (training data or misclassified bags) is being visualized}''; ``\textit{Provides additional information, maybe as a tooltip, indicating more information about instances in the instances projection space like type, instance's preview, etc. It would be excellent.}''

All participants successfully finished the user study within approximately 35 minutes. The majority of participants updated the prototypes of misclassified bags using the bags projection space. Only one explicitly used the instance space for this purpose.

The final accuracy achieved by the participants on the Corel-300 dataset is presented in Table \ref{tab:Corel300TestUsability}. As we can see, the average is 86.16\%, which is comparable to the 83.8\% achieved in Section \ref{subsec:casetwo}. This is because most participants selected a similar training set, learning from the instructions that it was possible to obtain good classification models by choosing bags located both on leaves and the core of the tree (see Section \ref{sec:ImpactBagsPositioning}). 

\begin{specialtable}[H] 
\caption{Results of multi-instance classification obtained by each participants on Corel-300 dataset.\label{tab:Corel300TestUsability}}
\begin{tabularx}{\linewidth}{>{\centering\arraybackslash}X>{\centering\arraybackslash}X>{\centering\arraybackslash}X>{\centering\arraybackslash}X>{\centering\arraybackslash}X>{\centering\arraybackslash}X>{\centering\arraybackslash}X}
\toprule
&\multicolumn{5}{c}{\textbf{Participants}} & \\
\cmidrule{2-6} 
\textbf{Task} &\textbf{User1} &\textbf{User2}&\textbf{User3}&\textbf{User4}&\textbf{User5}&\textbf{Average}\\
\midrule
Corel-300 &85.87&87.28&85.47&87.42&84.75&\textbf{86.16}\\
\bottomrule
\end{tabularx}
\end{specialtable}

\section{Statistical Analysis}
\label{sec:StatisticalAnalysis}

The statistical analysis was carried out using the main MIL benchmark datasets: Corel-1000, Corel-2000, Musk1, Musk2, Elephant, Fox and Tiger. The statistical non-parametric test proposed by~\cite{Garcia2010} was used to compare the results across the most relevant competing methods, which produces a ranking, as shown in Table~\ref{table:Friedman}, a contrast estimation table between the methods, as shown in Table~\ref{table:FriedmanContrast}, and post hoc tests as shown in Table~\ref{table:FriedmanPostHoc}, in order to verify which methods are significantly different from the best method, i.e., the top-ranked one.

Our MILTree-Med was considered by the Friedman test to be the first in the ranking (see Table~\ref{table:Friedman}) with $p\le0.01$; this is confirmed by the contrast estimation table, showing that it has higher (positive) average accuracy when compared with all other methods (see Table~\ref{table:FriedmanContrast}). According to the Holm's post hoc procedure (see Table~\ref{table:FriedmanPostHoc}), the MILTree-Med accuracy is significantly better than MI-SVM, EMDD, mi-SVM and DD-SVM, but it is not different when compared with MILES, MILDE and MILTree-SI. Li's post hoc procedure found that only MILDE is comparable with our MILTree-Med approach, while the remaining ones present statistically lower accuracies (see Table~\ref{table:FriedmanPostHoc}). These results corroborate our findings, encouraging the use of visual tools to support different MIL tasks.

\begin{specialtable}[H] 
\caption{Average rankings. Friedman statistic $23.5$ according to $\chi$-square with 7 degrees of freedom; $p=0.0014$.\label{table:Friedman}}
\begin{tabularx}{\linewidth}{>{\centering\arraybackslash}X>{\centering\arraybackslash}X}
\toprule
\textbf{Method}&\textbf{Ranking}\\
\midrule
MILTree-Med&2.3571\\
MILDE&3.0714\\
MILES&3.4285\\
MILTree-SI&3.5714\\
DD-SVM&4.5714\\
mi-SVM&5.5714\\
EMDD&6.2142\\
MI-SVM&7.2142\\
\bottomrule
\end{tabularx}
\end{specialtable}\vspace{-6pt}

\begin{specialtable}[H] 
\caption{Contrast Estimation between the methods on each row with respect to the methods on each column, considering different datasets. Positive values indicate that the method in the row presented higher average accuracy than the method in the column. The proposed methods are MILTree-Med (MT-Med) and MILTree-SI (MT-SI)\label{table:FriedmanContrast}}
\tablesize{\footnotesize}
\begin{tabularx}{\linewidth}{cc>{\centering\arraybackslash}Xcc>{\centering\arraybackslash}Xc>{\centering\arraybackslash}X>{\centering\arraybackslash}X}
\toprule
 &\textbf{MT-Med}&\textbf{MT-SI}&\textbf{MI-SVM}&\textbf{mi-SVM}&\textbf{EMDD}&\textbf{DD-SVM}&\textbf{MILES}&\textbf{MILDE}\\
\midrule
MT-Med&0.000&1.112&12.94&8.693&12.05&5.450&4.530&2.825\\
\midrule
MT-SI&$-$1.112&0.000&11.83&7.580&10.94&4.338&3.418&1.713\\
\midrule
MI-SVM&$-$12.94&$-$11.83&0.000&$-$4.250&$-$0.895&$-$7.492&$-$8.412&$-$10.12\\
\midrule
mi-SVM&$-$8.693&$-$7.580&4.250&0.000&3.355&$-$3.242&$-$4.163&$-$5.867\\
\midrule
EMDD&$-$12.05&$-$10.94&0.895&$-$3.355&0.000&$-$6.597&$-$7.518&$-$9.222\\
\midrule
DD-SVM&$-$5.450&$-$4.338&7.492&3.242&6.597&0.000&$-$0.920&$-$2.625\\
\midrule
MILES&$-$4.530&$-$3.418&8.412&4.163&7.518&0.920&0.000&$-$1.705\\
\midrule
MILDE&$-$2.825&$-$1.713&10.12&5.867&9.222&2.625&1.705&0.000\\
\bottomrule
\end{tabularx}
\end{specialtable}\vspace{-6pt}

\begin{specialtable}[H] 
\caption{Holm's  and Li's test results for $\alpha=0.05$ (Friedman). Holm's procedure rejects those hypotheses for $p\le0.01$; Li's procedure rejects those hypotheses for $p \le 0.022$.\label{table:FriedmanPostHoc}}%This table has not been referred to within the text of the manuscript. As tables and figures must be mentioned in the main text, please check and revise.
%Authors: Fixed. The table has been referred to at the start of section 8.
\begin{tabularx}{\linewidth}{>{\centering\arraybackslash}X>{\centering\arraybackslash}X>{\centering\arraybackslash}X>{\centering\arraybackslash}X}
\toprule
\boldmath{$i$}&\textbf{Method}&\textbf{Holm}&\textbf{Li}\\
\midrule
7&MI-SVM    &\textbf{0.001}&\textbf{0.022}\\
6&EMDD      &\textbf{0.001}&\textbf{0.022}\\
5&mi-SVM    &\textbf{0.001}&\textbf{0.022}\\
4&DD-SVM    &\textbf{0.010}&\textbf{0.022}\\
3&MILTree-SI&0.354&\textbf{0.022}\\
2&MILES     &0.413&\textbf{0.022}\\
1&MILDE     &0.585&0.050\\
\bottomrule
\end{tabularx}
\end{specialtable}

\section{Conclusions}
\label{sec:conclusions}
 In this paper, we propose MILTree for visual data mining in multiple-instance learning scenarios using an intuitive two-level tree structure that resembles MIL data models. While visually supporting data understanding, our approach also handles multiclass problems: the MILTree-SI selection method aims to uncover the most representative instances in both positive and negative bags, where negative bags could also have positive instances; the MILTree-Med method uses a clustering algorithm to partition unlabeled instances in search of positive and negative clusters to identify adequate instance prototypes. 

Besides producing comparable or better accuracy with respect to state-of-the-art methods, our MILTree-based techniques allow the user to take part in every step of the multiple-instance classification process, such as data exploration, sampling for training, model updating (both automatic and manual) and validating the classification model. 

The method has been tested on datasets of various sizes, and users have found positive aspects of the approach, as well as limitations, mainly related to interactive functions in the current prototype system. 

Our methods and techniques combined are, to the best of our knowledge, the first complete set of visual tools to support MIL learning.

Because we deal with data that are organized in multiple levels (bag and instance levels in the case of MIL), future work can explore related tasks, such as hierarchical clustering or learning from label proportions \cite{Yu2014, Stolpe2011}, in which the data are organized in groups that can be viewed as bags, and only the proportion of each class in each bag is known. An alternative high precision way of organizing the samples in bags might be to use multidimensional projections. While we would lose the hierarchical organization, there might be benefits in the precision of the display. This is a venue worth pursuing. A visual alternative for very large datasets is also an expected development from this work.

\vspace{6pt} 
\authorcontributions{Conceptualization, methodology, formal analysis and writing: S.C., M.A.P. and R.M.; software and adaptation of visualization, S.C.; supervision and project administration, M.A.P. and R.M.; funding acquisition, R.M. All authors have read and agreed to the published version of the manuscript.}

\funding{This research was funded by FAPESP grant 2013/25055-2 and CNPq grant 134238/2013-3. M.A.Ponti was funded in part by FAPESP (grant \#2019/07316-0) and the CNPq fellowship 304266/2020-5. R. Minghim was funded in part by the CNPq fellowship 307411/2016-8.}
%MDPI: Please check the accuracy of funding data and any other information carefully.
%Authors: Everything is correct. Thanks.

%\institutionalreview{Not applicable.}

%\informedconsent{Not applicable.}

\dataavailability{The source code, data and/or other artifacts have been made available at \url{https://github.com/soniacq/MILTree}.} 

\conflictsofinterest{The authors declare no conflict of interest. The funders had no role in the design of the study; in the collection, analyses or interpretation of data; in the writing of the manuscript; or in the decision to publish the results.}

\end{paracol}
\reftitle{References}

\end{document}